\documentclass[11pt]{article}
\usepackage{hyperref}
\usepackage{amsmath, amsfonts, amssymb, mathrsfs}
\usepackage{dcolumn}
\usepackage[hypcap=false]{caption}
\usepackage{subcaption}
\usepackage{filemod}
\usepackage{natbib}
\usepackage[ruled, vlined]{algorithm2e}
\usepackage{floatrow}
\usepackage{setspace}
\usepackage{verbatim}
\usepackage{graphicx}
\usepackage{textcomp}
\usepackage{bbm}

\DeclareMathOperator*{\argmax}{argmax}

\usepackage{xcolor} 
\hypersetup{
colorlinks=true,
linkcolor=purple, 
urlcolor=black,
citecolor=purple, 
}

\AtBeginDocument{\colorlet{blue}{black}}

\title{\vspace{-1cm} Profile Bayesian Optimization for Expensive Computer Experiments}
\author{Courtney Kyger \thanks{Corresponding Author: Department of Statistics, 
	Virginia Tech, {\tt ston2874@vt.edu}} 
\and James Fernandez \thanks{Department of Mechanical and Aerospace 
	Engineering, NC State University}
\and John A. Grunenwald \footnotemark[2]
\and James Braun \footnotemark[2]
\and Annie S. Booth \thanks{Department of Statistics, Virginia Tech}}

\begin{document}

\maketitle

\begin{abstract} 
	We propose a novel Bayesian optimization (BO) procedure aimed at 
	identifying the ``profile optima'' of a deterministic black-box computer 
	simulation that has a single control parameter and multiple nuisance 
	parameters. The profile optima capture the optimal response values as a 
	function of the control parameter. Our objective is to identify 
	\textcolor{blue}{these optima} across the entire 
	plausible range of the control parameter. Classic BO, which targets a 
	single optimum over all parameters, does not explore the 
	entire control parameter range. Instead, we develop a novel two-stage 
	acquisition scheme to balance exploration across the control parameter and 
	exploitation of the profile optima, leveraging deep and shallow 
	Gaussian process surrogates to facilitate uncertainty 
	quantification. We are motivated by a computer 
	simulation of a diffuser in a rotating detonation combustion engine, which 
	returns the energy lost through diffusion as a function of 
	various design parameters. We aim to identify the lowest
	possible energy loss as a function of the diffuser's length; understanding 
	this relationship will enable well-informed design choices. Our 
	``profile Bayesian optimization''
	procedure outperforms traditional BO and profile optimization methods on a 
	variety of benchmarks and proves effective in our motivating application 
	\textcolor{blue}{against state-of-the-art multi-objective optimization}.
\end{abstract}

\noindent \textbf{Keywords:} expected improvement, 
Gaussian process, deep Gaussian process, surrogate, triangulation candidates, 
uncertainty quantification

\section{Introduction}\label{sec:intro}

\textcolor{blue}{Computer simulations} are becoming increasingly relevant in 
physical and engineering sciences, where they may supplement physical 
experimentation that is costly, dangerous, or infeasible. In this work we 
tackle profile optimization of these black-box computer simulations, seeking 
the optimal response value over the full support of a specified ``control 
parameter.'' Profile optimization can provide crucial information about 
trade-offs between performance and an important parameter, thereby supporting 
informed decision making.

\begin{figure}[H]
	\centering
	\includegraphics[width=.35\linewidth]{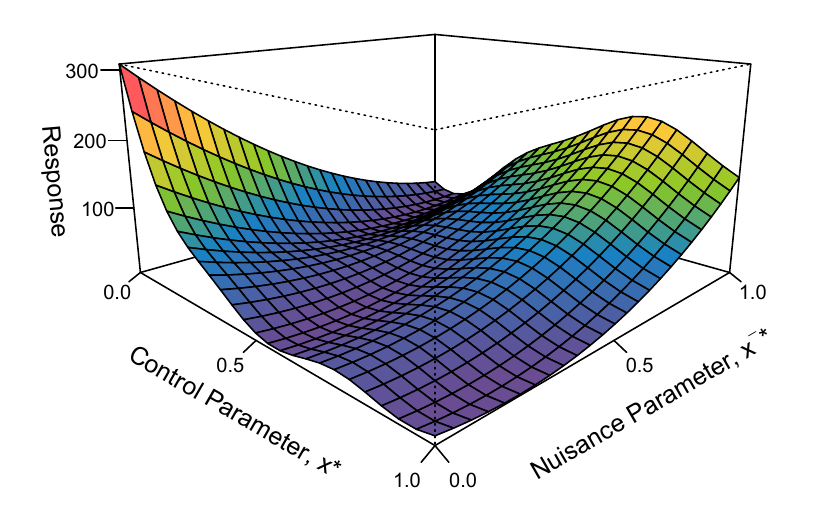}
	\includegraphics[width=.25\linewidth]{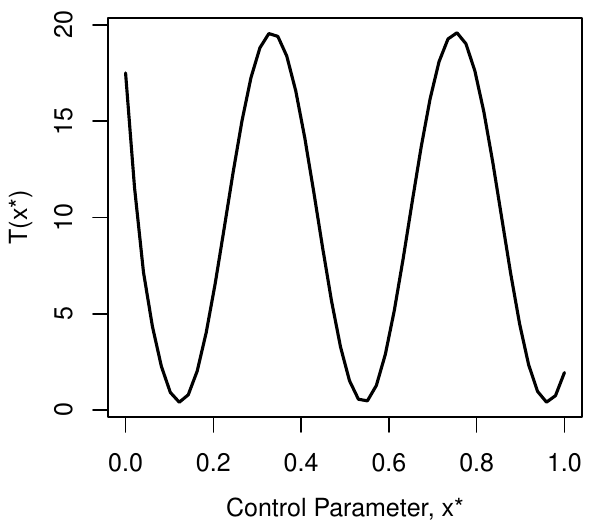} \hspace{12pt}
	\includegraphics[width=.25\linewidth]{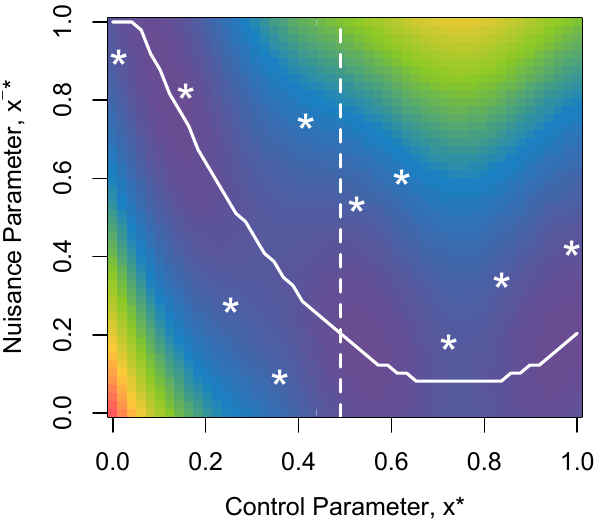}
	\caption{Branin function with one control parameter and one nuisance 
		parameter (left), with the true $T(x^\star)$ (center). Right panel 
		shows the location of $T(x^\star)$ in $\mathcal{X}$ (solid white), an 
		LHS design of size $n=10$ (white circles), and the vertical slice that 
		features later in Figure~\ref{fig:braninpost} (dashed white).}
	\label{fig:branin2D}
\end{figure}

Let $f:\mathcal{X}\rightarrow\mathbb{R}$ denote a black-box simulation that 
accepts $d$-dimensional input $\mathbf{x}\in\mathcal{X}\subset\mathbb{R}^d$ and 
returns scalar $y = f(\mathbf{x})$. Inputs are further delineated into a single 
control parameter $x^\star\in\mathcal{X}^\star\subset\mathbb{R}$ and nuisance 
parameters $\mathbf{x}^{-\star}\in\mathcal{X}^{-\star}\subset\mathbb{R}^{d-1}$ 
such that $\mathbf{x} = [x^\star, \mathbf{x}^{-\star}]$, 
$\mathcal{X} = \{\mathcal{X}^\star, \mathcal{X}^{-\star}\}$, and 
$f(\mathbf{x}) = f(x^\star, \mathbf{x}^{-\star})$. For a given control 
parameter, the profile optima of $f$ is defined as

\begin{equation}
	T(x^\star) := \min_{\mathbf{x}^{-\star}\in \mathcal{X}^{-\star}}
	f(x^\star, \mathbf{x}^{-\star}) \;\;\forall\;\; x^\star \in\mathcal{X}^\star,
	\label{eq:definetx}
\end{equation}
providing the minimum response that can be obtained for any value of the 
control parameter. To visualize this in a simple setting, Figure~\ref{fig:branin2D} 
portrays the two-dimensional Branin function 
\citep{surjanovic2013virtual} scaled to $\mathcal{X} = [0,1]^2$ with the 
resulting $T(x^\star)$.

Our real-world motivation stems from the design of a jet engine: a rotating 
detonation combustor (RDC) that pushes hot air through a diffuser to power a  
turbine. Computational fluid dynamics simulations compute the flow through the
diffuser, estimating the energy lost through diffusion as a function of the 
design parameters (i.e., curvature, bypass locations, and diffuser length).
\textcolor{blue}{Our primary focus is on diffuser length, the ``control 
parameter'' in this problem. Longer diffusers present the smallest energy 
loss, yet shorter diffusers are more cost-effective and practical in the 
design of the full engine. The remaining ``nuisance'' parameters, four 
curvature control points and the bypass location illustrated later in 
Figure~\ref{fig:diffuser}, have negligible effects on the cost and practicality, 
but still affect the energy loss. Thus, our objective is to estimate energy 
loss as a function of diffuser length, optimizing over our five nuisance 
parameters, to facilitate effective decision-making concerning the diffuser's design.}

\textcolor{blue}{Existing approaches in the field leverage multi-objective
optimization algorithms, considering the diffuser length as an objective itself
(though it is a design parameter) and seeking to jointly minimize diffuser 
length and energy loss \citep{braun2021aerothermal,verstraete2010design}.  
This approach is a mismatch to the actual objective (Eq.~\ref{eq:definetx})
and fails to provide quantification of uncertainty.  In contrast, profile
Bayesian optimization purposefully incorporates uncertainty quantification (UQ) and
addresses the entire range of the control parameter,
enabling more informed design choices. In Section \ref{sec:diffuser}, we 
compare our profile optimization 
results to a state-of-the-art multi-objective optimization of the RDC simulation,
highlighting the advantages of profile optimization.}

The key challenge in estimating $T(x^\star)$ is the computational expense of 
evaluating the computer \textcolor{blue}{experiment}. Simplified versions of 
our RDC simulation require 10 minutes of computation to produce a single 
observation, even with high performance parallel computing. Higher fidelity 
versions, \textcolor{blue}{which are still under development,} require several 
hours.

When simulation data is limited, as \textcolor{blue}{it is} in our motivating 
application, a surrogate model is essential. Surrogates are trained on 
observations of the simulator in order to provide predictions and UQ
at unobserved inputs. Gaussian process (GP) surrogates are 
the ``go-to'' choice \citep{santner2003design, rasmussen2006gaussian, 
gramacy2020surrogates}, but in their canonical form they may be too rigid to 
accommodate the complexity of modern computer simulations for physical systems. 
\textcolor{blue}{Notably, our motivating example is nonstationary in its 
response due to the combustion process.} Deep Gaussian processes 
\citep[DGPs;][]{damianou2013deep} offer increased flexibility by warping the 
input space through latent Gaussian layers and have been gaining significant 
attention as surrogates for nonstationary simulations 
\citep[e.g.,][]{rajaram2021empirical, marmin2022deep, sauer2023deep, 
yazdi2024deep}. Crucially, our contribution is suitable for both GP and DGP 
surrogates. 

Direct optimization of expensive computer experiments, i.e., 
$\min_{\mathbf{x}\in\mathcal{X}} f(\mathbf{x})$, is a well studied problem, 
usually tackled through Bayesian optimization \citep[BO;][]{frazier2018bayesian,
pourmohamad2021bayesian,wang2023recent}. BO involves a greedy feedback loop 
where design points are sequentially selected based on information from a 
surrogate model that has been trained on the previously observed data and 
is updated after each new acquisition. Common acquisition strategies 
include maximization of expected improvement \citep[EI;][]{jones1998efficient,
zhan2020expected} and Thompson sampling \citep{thompson1933likelihood}.
Naturally, BO methods that target the global optimum are 
ill-equipped for profile optimization as they are not designed to explore 
the entire control parameter space.

\citet{ginsbourger2013profile} proposed ``profile expected improvement'' (PEI) 
to remedy this disconnect. \textcolor{blue}{PEI, as we will review in 
Section~\ref{sec:pei}, modifies traditional EI to target profile optima 
rather than the traditional global optimum. One unsolved challenge of the 
original methology proposed by \citet{ginsbourger2013profile} is the 
computational burden of optimization over dense grids, as the requisite 
computations are incompatible with numerical optimizers.} We suspect this is 
the reason they only provided results on two-dimensional problems. 
\textcolor{blue}{Even when we adapt the implementation to facilitate faster 
acquisitions (more on this in Section~\ref{sec:profilebo}), the tendency to 
under-explore and over-exploit, a well-known limitation of traditional EI} 
\citep{qin2017improving, berk2018exploration, chen2024hierarchical} is 
inherited by PEI, as we illustrate in our simulation studies in Section 
\ref{sec:toyproblems}.

We propose a novel profile Bayesian optimization (PBO) procedure to effectively 
estimate $T(x^\star)$ with limited observations of the black-box function 
$f(\mathbf{x})$. \textcolor{blue}{First, we leverage joint posterior draws from 
a Bayesian GP or DGP surrogate to estimate $T(x^\star)$ with thorough 
uncertainty quantification. The next acquisition, $\mathbf{x}_{n+1}$, is 
selected in two stages: first, choosing $x^\star_{n+1}$ based on exploration of 
$\mathcal{X}^\star$, then choosing $\mathbf{x}^{-\star}_{n+1}$ based on 
exploitation of $T(x^\star_{n+1})$ through optimization of PEI along the 
$x^\star_{n+1}$ slice. The intentional exploration of the first step provides 
effective coverage of the control parameter space, while the exploitation of 
the second step hones in on the profile optima. To alleviate computational 
bottlenecks in our procedure, we also propose a strategic adaptation of 
triangulation candidates \citep[``tricands'';][]{gramacy2022triangulation}.}

The remainder of this paper is organized as follows. Section~\ref{sec:review} 
reviews the essential building blocks of our method. 
Section~\ref{sec:profilebo} details our PBO procedure. 
Section~\ref{sec:toyproblems} provides a variety of synthetic examples to 
validate our approach against state-of-the-art alternatives. Finally, we 
present the results of our method on the RDC in Section~\ref{sec:diffuser} and 
conclude in Section~\ref{sec:summary}.

\section{Building Blocks}\label{sec:review}

In this section, we detail the building blocks necessary for our contribution, 
including GP and DGP surrogates, traditional expected improvement, and profile 
expected improvement. Throughout, we use lowercase letters to represent 
scalars, bold lowercase letters to represent vectors, and bold uppercase 
letters to represent matrices. Specifically, let $\mathbf{X}$ denote a matrix 
of row-stacked inputs with response $\mathbf{y} = f(\mathbf{X})$. In all 
exercises, we prescale inputs to the unit hypercube ($\mathcal{X} = [0,1]^d$) 
and prescale responses to zero mean with unit variance before fitting our 
surrogates.

\subsection{GP and DGP Surrogates}\label{ss:gp}

A Gaussian process prior assumes responses were generated as a realization of a 
random Gaussian process such that any finite set of observations is distributed 
as a multivariate normal distribution, e.g., $\mathbf{y} \sim \mathcal{N} 
\left(\boldsymbol\mu,\, K(\mathbf{X})\right)$. After centering responses, we set $\boldsymbol\mu =\mathbf{0}$. We define the covariance matrix 
$K(\mathbf{X})$ with elements $K(\mathbf{X})^{(ij)} = \tau^2 
\left(k(||\mathbf{x}_i - \mathbf{x}_j||^2) + g\mathbb{I}_{i=j}\right)$, where 
kernel $k(\cdot)$ returns the correlation between $y_i$ and $y_j$ as a 
function of the Euclidean distance between their inputs, $\mathbf{x}_i$ and
$\mathbf{x}_j$. For our deterministic computer simulations, we fix $g = 1\times 
10^{-6}$ to preserve interpolation while providing numerical stability. Scale 
parameter $\tau^2$ and any other hyperparameters within the kernel may be 
inferred through maximum likelihood estimation or sampled in a Bayesian 
framework.

\begin{figure}[H]
	\centering
	\includegraphics[width=0.32\linewidth]{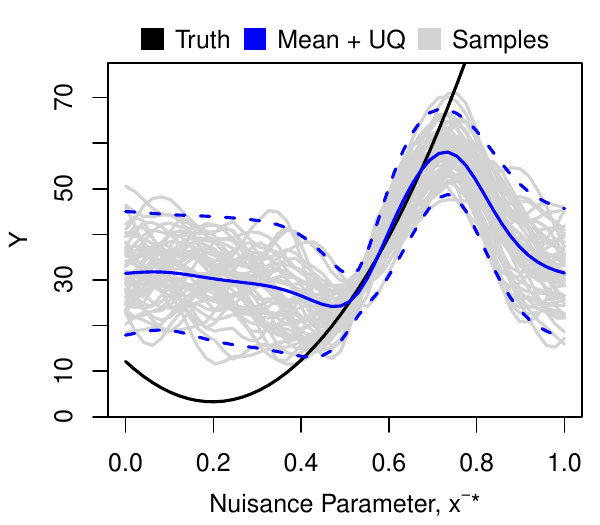}
	\includegraphics[width=0.32\linewidth]{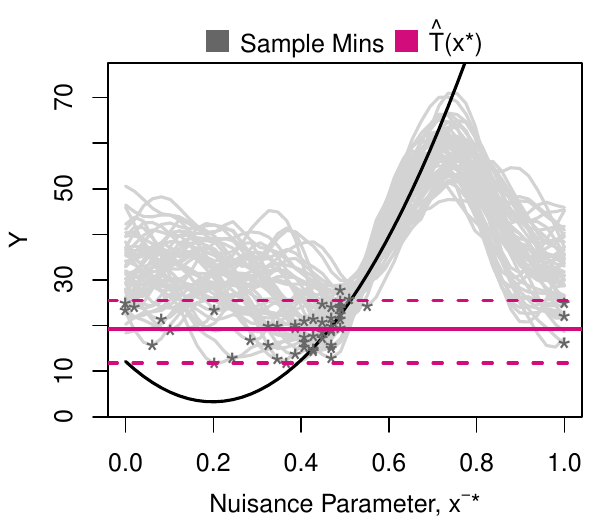}
	\includegraphics[width=0.32\linewidth]{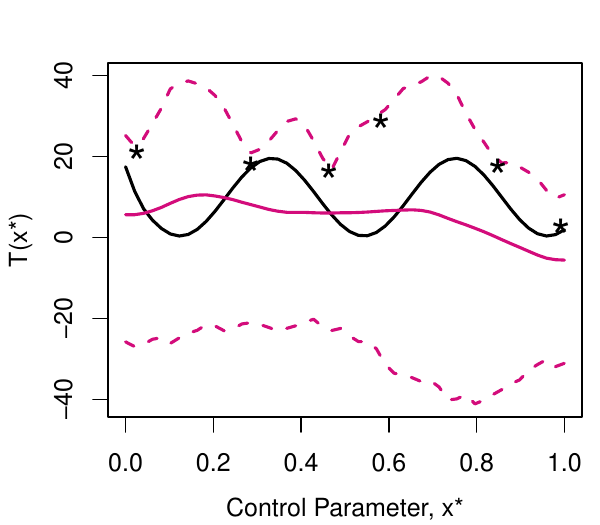}
	\caption{{\it Left:} GP trained on a random LHS of size $n=10$ (white circles in
	Figure~\ref{fig:branin2D}) for the 2D Branin function, shown along the slice $x^\star=0.48$ 
	with the posterior mean and 95\% CI in blue. Gray lines show 50 samples from this
	posterior.
	{\it Center:} The minimum values of each posterior sample (dark gray stars).
	Dark pink lines mark the the mean and 95\% quantiles of 1,000 such minimums.
	{\it Right:} Estimation of the full $\hat{T}(x^\star)$ (dark pink) with the 
	truth in solid black.
	Black stars mark observed $(x^\star_i, y_i)$ for $i=1,\dots,n$ (some observations 
	are cut off above).}
	\label{fig:braninpost}
\end{figure}

Conditioned on $n$ observations, $\mathcal{D}_n = \{\mathbf{X}_n, \mathbf{y}_n\}$, the posterior 
distribution of $f$ is itself a Gaussian process. When restricted to a finite 
set of locations $\mathbf{X}_p$ of size $n_p\times d$, the posterior is 
expressed as
\begin{equation}\label{eq:gppost}
	f_n(\mathbf{X}_p)\mid \mathcal{D}_n \sim \mathcal{N}_{n_p} 
	\left(\mu_n(\mathbf{X}_p),\, \Sigma_n(\mathbf{X}_p)\right)
	\;\;\textrm{where}\;\;\;
	\begin{aligned}
		\mu_n(\mathbf{X}_p) &= K(\mathbf{X}_p, \mathbf{X}_n)K(\mathbf{X}_n)^{-1}\mathbf{y}_n \\
		\Sigma_n(\mathbf{X}_p) &= K(\mathbf{X}_p) - 
		K(\mathbf{X}_p, \mathbf{X}_n)K(\mathbf{X}_n)^{-1}K(\mathbf{X}_n, \mathbf{X}_p).
	\end{aligned}
\end{equation}
Here, $K(\mathbf{X}_p, \mathbf{X}_n)$ denotes the $n_p\times n$ matrix of 
covariances between each row of $\mathbf{X}_p$ and each row of $\mathbf{X}_n$. 
To demonstrate, we trained a GP surrogate for the Branin function using a 
random Latin hypercube sample \citep[LHS;][]{mckay2000comparison} of size 
$n=10$ (indicated by the white circles in Figure~\ref{fig:branin2D}). The left 
panel of Figure~\ref{fig:braninpost} shows $\mu_{10}(\mathbf{X}_p)$ (solid 
blue) and its 95\% credible interval (dashed blue) along the slice $x^\star = 
0.48$, where $\mathbf{X}_p$ contains a fine and evenly-spaced grid of 
$x^{-\star}$ values in one dimension. The GP offers nonlinear predictions with 
effective UQ. Posterior uncertainty is lower near the training data—a GP 
hallmark. 

Joint posterior samples may be obtained by sampling realizations from this 
multivariate normal distribution, such as those shown in gray in 
Figure~\ref{fig:braninpost}. The granularity of these joint samples depends on 
the density of the predictive locations, $\mathbf{X}_p$. A fine grid, while 
sufficient in \textcolor{blue}{one or two dimensions, becomes too cumbersome in 
higher dimensions, such as our motivating 6D example.} Even though the 
computation of $\mu_n(\mathbf{X}_p)$ and $\Sigma_n(\mathbf{X}_p)$ in 
Eq.~(\ref{eq:gppost}) is at most quadratic in $n_p$, drawing joint samples from 
$f_n(\mathbf{X}_p)$ requires decomposition of $\Sigma_n(\mathbf{X}_p)$, which 
is cubic in $n_p$. To avoid this bottleneck, it is common to stick with 
point-wise predictions, treating each row of $\mathbf{X}_p$ independently. But 
when joint posterior samples are necessary (as in our proposed framework), 
Vecchia approximation \citep{vecchia1988estimation} can offer some reprieve. 
Vecchia-approximated joint posterior samples are obtained by sequentially 
drawing observations from the univariate Gaussian posterior distribution of 
each predictive location, while allowing these posteriors to condition on 
previously sampled values. Moving forward, we will use Vecchia approximation 
for faster posterior sampling; see \citet{katzfuss2020vecchia} and 
\citet{katzfuss2021general} for further details on Vecchia-approximated GPs.

While GPs offer flexible nonlinear regression, they are limited by the 
stationarity of the covariance kernel. The GP covariance structure is based 
solely on the Euclidean distance between inputs, which can be a severe handicap
for response surfaces with regime shifts or starkly varying dynamics
\citep{booth2024non}. Deep Gaussian processes \citep{damianou2013deep} upgrade 
typical GPs through functional compositions, providing additional flexibility
to accommodate complex nonstationary functions. Latent Gaussian layers serve as 
spatial deformations of the input space \citep{sampson1992nonparametric, 
schmidt2003bayesian}. While they may vary in width and depth, the most common 
structure for DGP surrogates contains one latent layer with conditionally 
independent ``nodes'' of dimension $d$ \citep{sauer2023active}. Specifically,
\begin{equation}\label{eq:dgp}
	\begin{aligned}
		\mathbf{y} &\sim \mathcal{N}\left(\boldsymbol\mu_y,\, K(\mathbf{W})\right) \\
		\mathbf{w}_i &\sim \mathcal{N}\left(\boldsymbol\mu_w,\, K(\mathbf{X})\right)
	\end{aligned}
	\quad\textrm{for}\quad i=1,\dots,d
	\quad\textrm{where}\quad 
	\mathbf{W} = \begin{bmatrix} \mathbf{w}_1 & \dots & \mathbf{w}_d \end{bmatrix}.
\end{equation}

Prior means of $\boldsymbol\mu_y = \boldsymbol\mu_w = \mathbf{0}$ are common,
although setting the prior mean of $\mathbf{w}_i$ to the $i^\textrm{th}$ column
of $\mathbf{X}$ may provide a helpful degree of regularization. Latent 
$\mathbf{W}$ is the driver of nonstationary flexibility but can be difficult to 
infer. Bayesian posterior integration through elliptical slice sampling 
\citep[ESS;][]{murray2010elliptical} has been shown to excel in modeling of 
complex deterministic computer \textcolor{blue}{simulations} 
\citep[e.g.,][]{sauer2023non,ming2023deep, booth2025contour}. Conditioned on 
$\mathcal{D}_n$, ESS provides posterior samples $\mathbf{W}_n^{(t)}$, where $t$ 
indexes the MCMC iteration. These in turn facilitate inference of warped 
$\mathbf{W}_p^{(t)}$ for predictive locations $\mathbf{X}_p$. Ultimately, joint 
posterior draws of $f_n(\mathbf{X}_p)$ may be obtained through application of 
Eq.~(\ref{eq:gppost}) with warped/inferred $\mathbf{W}_n^{(t)}$ and 
$\mathbf{W}_p^{(t)}$ in place of $\mathbf{X}_n$ and $\mathbf{X}_p$. See 
\citet{sauer2023active} for specifics. Vecchia approximation is also available 
to enable faster DGP posterior sampling \citep{sauer2023vecchia}. While we use 
traditional GPs for our illustrative examples 
(Figures~\ref{fig:branin2D}-\ref{fig:braninexampleei}), the nonstationary 
flexibility of the DGP is essential for our motivating RDC simulation.

With a trained GP or DGP surrogate in-hand, estimation of $T(x^\star)$ may 
proceed as follows. First, generate joint samples from the posterior 
distribution at $\mathbf{X}_p\in\mathcal{X}$ locations (we will discuss the 
choice of $\mathbf{X}_p$ in Section~\ref{sec:profilebo}). Then, save the 
minimum response value attained in each posterior sample for each unique value 
of $x^\star\in\mathbf{X}_p$. Together, the empirical distribution of these 
minimums serves as the estimate $\hat{T}(x^\star)$.  We will summarize this 
distribution through its mean, denoted $\mu_T(x^\star)$, and $95^\textrm{th}$ 
percentiles, denoted $\mathrm{CI}_T(x^\star)$. Grabbing the minimum from a 
joint posterior sample is reminiscent of Thompson sampling, where the minimum 
of a single sample is used as the next acquisition in a BO loop.

Figure~\ref{fig:braninpost} illustrates this process for the GP trained on 10 
observations of the Branin function. \textcolor{blue}{We drew 1,000 joint 
posterior samples at each input location from a dense two-dimensional grid of 
size 10,000 ($\mathbf{X}_p = [0, 0.01, \dots, 0.99, 1]\times [0, 0.01, \dots, 
0.99, 1]$).} The left panel shows 50 of these samples (gray lines) along the 
slice $x^\star=0.48$. For each sample, we isolate the minimum observed 
$y$-value; these are marked by the dark gray stars in the center panel. 
Although we only show 50 here for demonstration, we use the minimums of all 
1,000 samples to generate the mean and 95th percentiles shown in solid/dashed 
pink. We repeat this process of grabbing and aggregating the minimums along 
each $x^\star$ slice, using the same joint posterior samples (which were drawn 
over the entire two-dimensional space). Sampling from the posterior over the 
full $d$-dimensional space is essential to incorporate the smoothly varying 
covariance structure across $x^\star$. Together, the mean/percentiles taken 
along each slice provide an estimate of the complete $T(x^\star)$, as shown in 
the right panel. \textcolor{blue}{The overly smooth $\hat{T}(x^\star)$ and its 
large uncertainty are due to the limited training data ($n=10$ in 2D).  We can 
reduce uncertainty and improve estimation through strategic sequential 
acquisitions.}

\subsection{Expected Improvement}

Given a surrogate $f_n(\mathbf{x})$, trained on observed $\mathcal{D}_n$, 
Bayesian optimization selects the next input as $\mathbf{x}_{n+1} = 
{\argmax}_{\mathbf{x}\in\mathcal{X}}\; h(\mathbf{x})$ where $h(\mathbf{x})$ is 
some ``acquisition function'' that quantifies the utility of a potential 
acquisition. If the acquisition function has a closed-form and is quick to 
evaluate, it may be fed through a numerical optimizer, such as a multi-start 
gradient descent. If evaluation of $h(\mathbf{x})$ is cumbersome, a discrete 
search over a set of potential candidates is usually preferred. Once selected, 
$y_{n+1} = f(\mathbf{x}_{n+1})$ is observed, $n$ is incremented, the surrogate 
is updated, and the process is repeated until a stopping criterion is met or 
the computation budget is exhausted.

For typical BO, the most common acquisition function is expected improvement
\citep{jones1998efficient}. For global optimization, a new acquisition would 
offer improvement if its response value is below the lowest response we have 
observed thus far (denoted $y_\mathrm{min}$). EI takes the expectation of this 
improvement, which has a closed form under a GP surrogate:
\begin{equation}
	\mathrm{EI}\left(\mathbf{x}\mid\mathcal{D}_n\right) = 
	\left(\textcolor{blue}{y_\textrm{min} - \mu_n(\mathbf{x})}\right)
	\Phi\left(\frac{\textcolor{blue}{y_\textrm{min} - \mu_n(\mathbf{x})}}{\sigma_n(\mathbf{x})}\right) + 
	\sigma_n(\mathbf{x})
	\phi\left(\frac{\textcolor{blue}{y_\textrm{min} - \mu_n(\mathbf{x})}}{\sigma_n(\mathbf{x})}\right) 
	\textrm{  where  } y_\mathrm{min} = \min(\mathbf{y}_n).
	\label{eq:tradei}
\end{equation} 
Here, $\mu_n(\mathbf{x})$ and $\sigma_n(\mathbf{x}) = 
\sqrt{\Sigma_n(\mathbf{x})}$ follow Eq.~(\ref{eq:gppost}), $\Phi(\cdot)$ is the 
standard normal CDF, and $\phi(\cdot)$ is the standard normal PDF.  
\textcolor{blue}{Since this EI formula is merely contingent on a Gaussian 
posterior, it is similarly applicable to DGP surrogates whose posterior is 
conditionally Gaussian given latent $\mathbf{W}$.  DGP posterior moments 
conditioned on $\mathbf{W}^{(t)}$ are indexed by iteration $t$, resulting in 
$\mathrm{EI}^{(t)}$, with expectation then taken over $t$.}

\begin{figure}[H]
	\includegraphics[width=.3\linewidth]{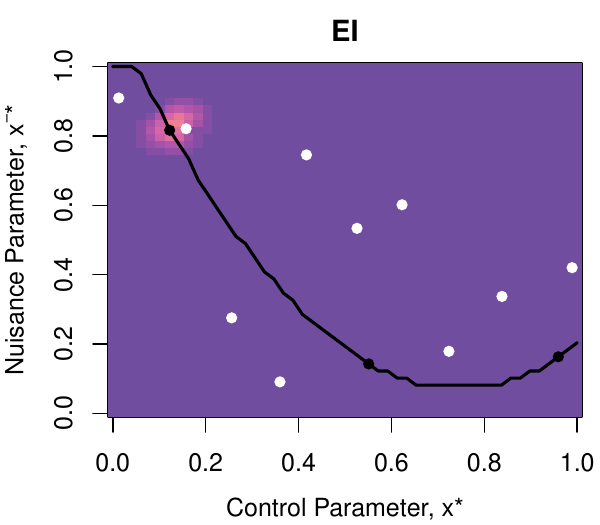}
	\includegraphics[width=.3\linewidth]{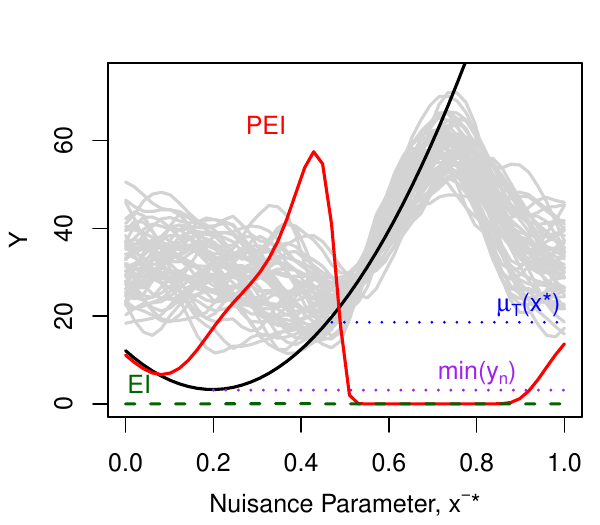}
	\includegraphics[width=.3\linewidth]{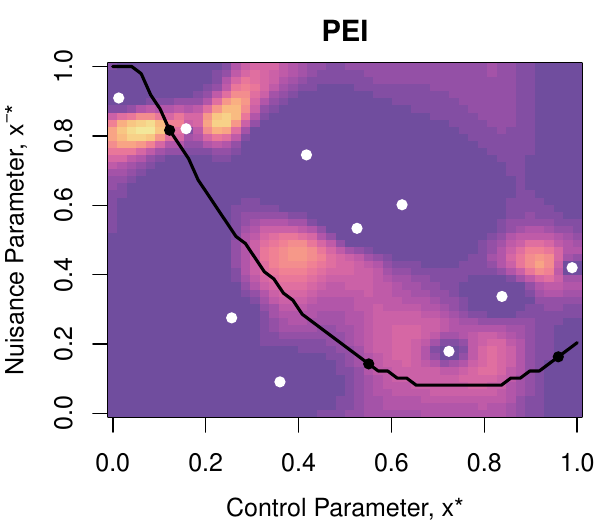}
	\caption{\textcolor{blue}{{\it Left/Right:} EI and PEI (purple/low, yellow/high) 
	from a GP surrogate trained on 10 observations of the 
	Branin function. White circles mark training data locations. Black 
	line/circles mark the true profile/global optima. {\it Center:} The same 
	slice shown in Figure \ref{fig:braninpost}, this time with EI (dashed 
	green) and PEI (solid red) overlayed. Horizontal dotted lines mark 
	$y_\textrm{min}$ (purple, used for EI) and estimated $\mu_{T}(x^\star)$ 
	(blue, used for PEI).}}
	\label{fig:braninexampleei}
\end{figure}

To demonstrate, the left panel of Figure~\ref{fig:braninexampleei} shows EI 
across $\mathcal{X}$ for our running Branin function example. The black circles 
mark the true global minima (there are three for this function). Notice, one of 
our random LHS points (white circles) landed very close to a true minimum. The 
resulting GP surrogate does not think there is improvement to be made 
elsewhere; the only region of high EI is in the locality of this observed low 
point. If our goal is to locate a global optimum, this EI surface would lead us 
right to it. But if our goal is to learn all the profile optima (the entire 
black line), EI is not an effective acquisition function. It is not designed to 
explore across the range of the control parameter. As we can see in the center
panel of Figure~\ref{fig:braninexampleei}, EI is near zero on this slice 
because $y_{min}$ is below the posterior samples (the gray lines) generated by 
our surrogate. \textcolor{blue}{These near-zero expected improvement values
would not facilitate an effective acquisition on this slice.}

\subsection{Profile Expected Improvement} \label{sec:pei}

In traditional optimization, ``improvement'' is only achieved if the acquired 
response value is lower than all previously observed responses, i.e., 
$f(\mathbf{x}_{n+1}) < y_\textrm{min}$. But in profile optimization, a new 
acquisition would offer ``improvement'' if its response is below the previously 
observed responses {\it for the same $x^\star$ value.} EI is not suited for 
profile optimization simply because of its use of $y_\textrm{min}$. 
\textcolor{blue}{For an $x^\star$ whose minimum lies above $y_\textrm{min}$, EI 
would provide a flat non-informative acquisition surface, as illustrated
in Figure \ref{fig:braninexampleei}.} The natural solution 
is to upgrade $y_\textrm{min}$ to a new quantity that acts as a function of 
$x^\star$, say $t_\textrm{min}(x^\star)$. This quantity should capture our 
existing knowledge about the lowest response value for a particular $x^\star$. 
Unfortunately, given our limited training data we can not rely on 
$\mathcal{D}_n$ to provide $t_\textrm{min}(x^\star)$ directly. Instead, we need 
an estimate that can be evaluated for any $x^\star\in\mathcal{X}^\star$ beyond 
those that are in $\mathcal{D}_n$.

\citet{ginsbourger2013profile} proposed using the estimated $\mu_{T}(x^\star)$, 
as described in Section~\ref{ss:gp}, in place of $y_\textrm{min}$ in the 
traditional EI criterion. They also implemented a strategic safeguard to 
protect against situations where $\mu_{T}(x^\star)$ may be low simply due to 
high uncertainty. We have already seen that very low $y_\textrm{min}$ values 
result in flat and uninformative EI surfaces 
(Figure~\ref{fig:braninexampleei}). To avoid this phenomenon with PEI, 
\citeauthor{ginsbourger2013profile} use the traditional $y_\textrm{min}$ value 
in areas where it is higher than $\mu_{T}(x^\star)$. Selecting the higher of 
these values is a conservative approach that avoids over-flattening the 
acquisition surface. Formally, this ``profile expected improvement'' criterion 
is defined as
\begin{equation}
	\mathrm{PEI}(\mathbf{x} = \{x^\star, \mathbf{x}^{-\star} \} \mid \mathcal{D}_n) =
		\left(\textcolor{blue}{t_\textrm{min}(x^\star) - \mu_n(\mathbf{x})}\right)
		\Phi \left(\frac{\textcolor{blue}{t_\textrm{min}(x^\star) - \mu_n(\mathbf{x})}}{\sigma_n(\mathbf{x})}\right) +
		\sigma_n(\mathbf{x}) \phi \left(\frac{\textcolor{blue}{t_\textrm{min}(x^\star) - \mu_n(\mathbf{x})}}{\sigma_n(\mathbf{x})}\right) \\
	\label{eq:pei}
\end{equation}
where $t_\textrm{min}(x^\star) = \max\left(y_\textrm{min}, 
\mu_{T}(x^\star)\right)$. \textcolor{blue}{Just as with EI, $\mu_n(\mathbf{x})$ 
and $\sigma_n(\mathbf{x}) = \sqrt{\Sigma_n(\mathbf{x})}$ can originate from either a
GP or DGP surrogate.}

We offer two visuals of PEI for our running Branin function example. First, the 
center panel of Figure~\ref{fig:braninexampleei} shows PEI along the slice 
$x^\star = 0.48$. The horizontal dotted blue line marks $\mu_{T}(x^\star = 
0.48)$. By using this higher value in place of $y_\textrm{min}$, PEI is able to 
offer an informative acquisition surface, targeting the minimum on this slice 
even though it is higher than the minimum response over the entire domain. As a 
second visual, the right panel of Figure~\ref{fig:braninexampleei} shows PEI 
across both dimensions. PEI targets more of the profile optima instead of 
focusing only on a global optimum. 

While PEI is a helpful extension of EI for profile optimization, its 
calculation is very burdensome. Optimization of PEI first requires estimation 
of $\mu_{T}(x^\star)$ for any potential $x^\star$. Estimation of 
$\mu_{T}(x^\star)$ requires potentially thousands of posterior samples at a 
dense set of predictive locations $\mathbf{X}_p$. These computations will not 
integrate nicely in a numerical optimizer. \citeauthor{ginsbourger2013profile} 
employed $50\times50$ grids in two dimensions but did not expand beyond a 
single control parameter and a single nuisance parameter. Furthermore, despite 
its attempts to target all profile optima, PEI tends towards excessive 
exploitation. Consider the right panel of Figure~\ref{fig:braninexampleei}; 
although there are local optima in the PEI surface across the range of 
$x^\star$, the global optimum (where $\mathbf{x}_\textrm{n+1}$ would be 
selected), still hovers around the location of $y_\textrm{min}$. 
\textcolor{blue}{As we will illustrate in Section~\ref{sec:toyproblems}, 
we find that PEI often clusters acquisitions 
around global optima and struggles to explore the entire $\mathcal{X}^\star$.}

\section{Profile Bayesian Optimization}\label{sec:profilebo}

Motivated by the drawbacks of PEI as an acquisition function for profile 
optimization, we propose a new profile Bayesian optimization procedure that 
involves two crucial upgrades. First, we develop a novel candidate scheme 
building upon the work of \citet{gramacy2022triangulation} that facilitates
efficient posterior sampling for the estimation of $T(x^\star)$ 
and enables acquisitions from smaller candidate sets. Second, we propose a 
two-stage acquisition method: the 
first step forces the exploration of $T(x^\star)$ across the full support of 
$x^\star$ by choosing the $x^\star$ with the largest uncertainty; the 
second step uses PEI to exploit the best acquisition in $\mathcal{X}^{-\star}$ 
for the chosen $x^\star$.

\subsection{Modified Tricands}

\textcolor{blue}{As discussed in Section~\ref{ss:gp},} estimation of $T(x^\star)$ 
requires joint random samples from the posterior distribution at a discrete 
number of input locations $\mathbf{X}_p$. The choice of $\mathbf{X}_p$ can 
make-or-break this estimation. Although these samples should be drawn over the 
entire $d$-dimensional space, consider first a slice along a particular 
$x^\star$, such as those shown in Figure~\ref{fig:braninpost}. For each sample 
(gray line), we grab the minimum response value observed at the $\mathbf{X}_p$ 
locations along that slice (dark gray stars). If $\mathbf{X}_p$ does not 
contain the point where the actual minimum is, then we will miss the minimum 
altogether. In one dimension, we can be pretty confident that a dense evenly 
spaced grid will include the minimum, but dense grids are infeasible in higher 
dimensions. Our RDC simulation has 6 inputs---even a semi-dense grid with 50 
points per dimension would require upwards of 15 billion points in 
$\mathbf{X}_p$.

Instead of grids, we leverage smaller ``candidate'' sets whose strategic 
allocation enables their size to be much less than the comparable grid. The 
number of points in $\mathbf{X}_p$ does not need to be large as long as it 
includes the locations where the minimums will be found. We can target these 
locations using geometric intuition about where extrema are likely to occur. In 
a GP surrogate, posterior uncertainty inflates as distance from the training 
data grows---notice $\Sigma_n(\mathbf{X}_p)$ in Eq.~(\ref{eq:gppost}) is only a 
function of the Euclidean distances among training and predictive locations. A 
random sample from the posterior is likely to have its lowest point in one of 
these regions of highest uncertainty---directly ``between'' training data 
locations. In one dimension, these locations include ``internal candidates'' at 
the midpoints of each consecutive pair of training locations and ``fringe 
candidates'' between the outermost observations and the boundary of 
$\mathcal{X}$.

\begin{figure}[H]
	\centering
	\includegraphics[width=.3\linewidth]{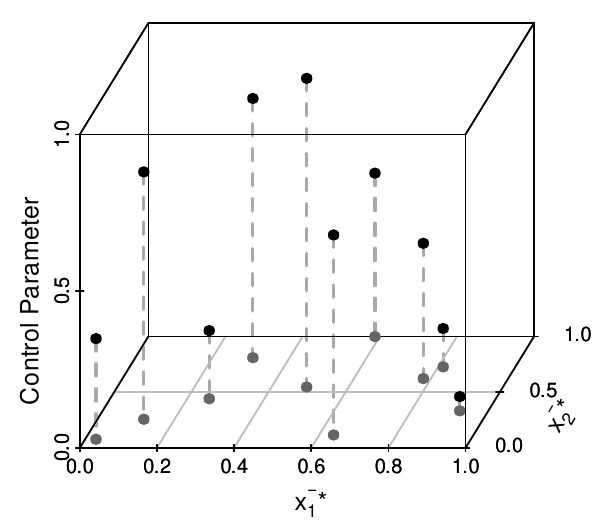}
	\includegraphics[width=.3\linewidth]{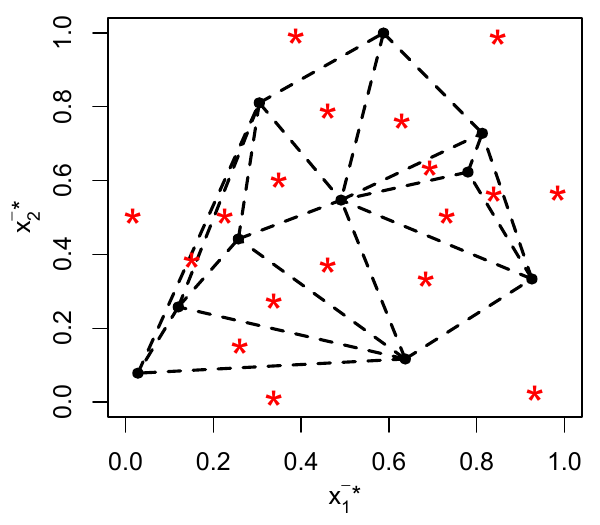}
	\includegraphics[width=.3\linewidth]{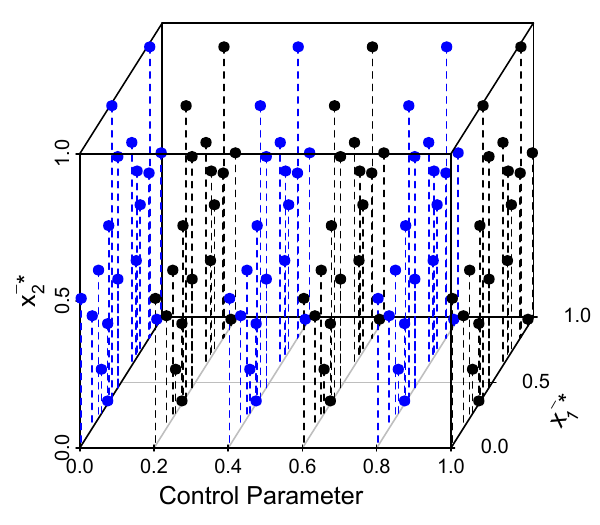}
	\caption{\textit{Left:} Projection of 3D LHS onto 2D nuisance 
	parameter space. \textit{Center:} Tricands (red stars) of 
	10-point LHS (black points). Dashed black lines represent the 
	Delaunay triangulation. 
	\textit{Right:} Tricands repeated on 6 slices for 
	$x^\star \in \{0,0.2,0.4,0.6,0.8,1\}$ (colors used for visual clarity).}
	\label{fig:tricands3D}
\end{figure}

In higher dimensions, the concept of ``between'' is a little murkier. 
\citet{gramacy2022triangulation} proposed ``tricands'' (short for triangulation 
candidates) for this very purpose. Tricands leverage a Delaunay triangulation 
of $\mathbf{X}_n$ to allocate candidates between existing training data 
locations. The triangulation connects observations in $\mathbf{X}_n$ creating 
non-overlapping triangles/tetrahedra/polychora/etc. depending on dimension. The 
dashed black lines in the center panel of Figure~\ref{fig:tricands3D} 
illustrate the Delaunay triangulation of some $\mathbf{X}_n$ (black circles) in 
two dimensions.

Tricands are allocated in two groups. Internal candidates are placed at the 
geometric centers of the triangulation's components. In 
Figure~\ref{fig:tricands3D}, these are the red stars inside the triangles. 
Fringe candidates are placed between the outermost edges and the boundaries of 
$\mathcal{X}$. In Figure~\ref{fig:tricands3D}, these are the red stars not 
contained in any triangle. \citet{gramacy2022triangulation} implemented a 
tuning parameter controlling how far to set fringe candidates between the 
triangulation and the border. In our exercises, we allocate fringe candidates 
at 90\% of the distance to the border, since we often find minima live near the 
boundary.

Through this strategic allocation conditioned on $\mathbf{X}_n$, tricands can 
cover regions where minima are likely to be sampled with a reasonably small 
candidate set size.  \textcolor{blue}{The number of tricands typically grows 
with the number of observed points and input dimension, see 
e.g., \citet[][Figure 3]{gramacy2022triangulation}, but it also depends on the 
locations of the observed points. In our synthetic and motivating examples, the 
number of tricands produced was significantly smaller than the equivalent 
$50^d$ grid. While our smaller-dimension examples entail modest candidate sets, 
for our higher-dimension examples, including the 6D RDC, we utilize random 
subsetting to limit the number of candidates in $\mathcal{X}^{-\star}$ to 
$100(d-1)$.}

Tricands provide an efficient set of candidates suitable for finding the 
minimum of a function. This is appropriate for our nuisance parameters, where 
the only objective is to find the minimum in $\mathcal{X}^{-\star}$ for each 
$x^\star$. Our objective regarding the control parameter is different: we seek 
to estimate $T(x^\star)$ across the full support of $x^\star$. Tricands does 
not offer the coverage we need over $\mathcal{X}^\star$, but this coverage can 
easily be achieved with a semi-dense grid. To fulfill both objectives, we 
combine a grid over the control parameter with tricands over the nuisance 
parameters.

We begin by projecting $\mathbf{X}_n$ onto the nuisance space, 
$\mathcal{X}^{-\star}$, to find $\mathbf{X}^{-\star}_n$ (i.e., $\mathbf{X}_n$ 
without the control parameter column). Then, we obtain tricands from 
$\mathbf{X}_n^{-\star}$, which we will denote $\{\mathbf{X}_\textrm{tri}\mid 
\mathbf{X}_n^{-\star}\}$. If there is a single nuisance parameter, the tricands 
are simply midpoints. Otherwise, they leverage the Delaunay triangulation 
discussed above. We then generate a one-dimensional grid of points, 
$\mathbf{x}^\star_{\textrm{grid}}$, to span the full support of the control 
parameter. Ultimately, we form the \textcolor{blue}{Cartesian} product 
between the grid and tricands to obtain the full $d$-dimensional candidate set, 
which we denote as $\mathbf{X}_{\textrm{tri}^+}$:
\begin{equation}
	\mathbf{X}_{\textrm{tri}^+} := \mathbf{x}^\star_{\textrm{grid}} \textcolor{blue}{\times} 
	\left\{\mathbf{X}_{\textrm{tri}}\mid \mathbf{X}^{-\star}_n\right\}.
	\label{eq:tricplus}
\end{equation}

Figure~\ref{fig:tricands3D} depicts this process visually. The left panel 
displays an initial LHS over the full $d$-dimensional space ($d=3$ for this 
example). These training observations are then projected over the control 
parameter into $d-1$ dimensions, creating $\textbf{X}^{-\star}_n$ as 
illustrated by the gray lines and gray points in this figure. We generate 
tricands from the 2D $\textbf{X}^{-\star}_n$, denoted $\textbf{X}_\textrm{tri}$ 
(red stars in the center panel). Then, we create a semi-dense grid of 50 points 
for $x^\star$. To offer a cleaner visual, Figure~\ref{fig:tricands3D} shows 
only a 6-point $\textbf{x}^\star_\textrm{grid}$. We then perform the 
\textcolor{blue}{Cartesian} product of $\textbf{X}_\textrm{tri}$ with 
$\textbf{x}^\star_\textrm{grid}$ to find $\textbf{X}_{\textrm{tri}^+}$, as 
shown in the right panel.

While we designed this candidate set construction for estimation of 
$T(x^\star)$, we find that it also works well as a discrete set of candidates 
for the next acquisition, $\mathbf{x}_{n+1}$. In our profile Bayesian 
optimization procedure, we will select subsequent acquisitions from our 
proposed $\mathbf{X}_{\textrm{tri}^+}$. 

\subsection{Acquisitions}

Our modified tricands enable the use of PEI in higher dimensions, as they avoid 
the computational barrier of dense grids. But PEI as a sole acquisition metric 
will often cluster acquisitions near the global minima rather than acquiring 
points across the full support of $x^\star$ \textcolor{blue}{(see 
Figure~\ref{fig:Tstar3D} for an illustration of this)}. To remedy this 
lack of exploration, we propose a new two-stage acquisition procedure that 
prioritizes exploration of $\mathcal{X}^\star$ while still leveraging the 
exploitative power of PEI. Our primary aim is to drive down uncertainty in our 
estimation of $T(x^\star)$ while retaining accuracy, within the constraints of 
a limited training budget.

Given a surrogate $f_n$, trained on observations $\mathcal{D}_n=\{\mathbf{X}_n, 
\mathbf{y}_n\}$, we seek to acquire $\mathbf{x}_{n+1} = \{x_{n+1}^\star, 
\mathbf{x}_{n+1}^{-\star}\}$ to maximize learning of $T(x^\star)$ 
(Eq.~\ref{eq:definetx}). We first generate $\mathbf{X}_{\textrm{tri}^+}$ 
following Eq.~(\ref{eq:tricplus}), but with one modification. Since we are  
going to repeat this procedure multiple times (until our training budget is 
exhausted), we replace the fixed evenly-spaced grid 
$\mathbf{x}_\textrm{grid}^\star$, which would remain constant throughout the 
entire optimization, with a 1D LHS $\mathbf{x}_\textrm{LHS}^\star$, which is 
re-randomized for each iteration. The LHS in 1D offers effective coverage of 
$\mathcal{X}^\star$, while providing variability in the exact $x^\star$ values 
that are considered. We then estimate $\hat{T}(x^\star)$ using these candidates 
as described in Section~\ref{ss:gp}.  Let $\mathrm{CI}_T^\textrm{wd}(x^\star)$ 
represent the width between the 97.5\% and 2.5\% quantiles \textcolor{blue}{of 
$\hat{T}(x^\star)$} for a particular $x^\star$, which serves as a measure of 
uncertainty in our estimate. For our next acquisition, we select the $x^\star$ 
value with the largest uncertainty:
\begin{equation}\label{eq:xstarnext}
x_{n+1}^\star = \argmax_{x^\star\in\mathbf{x}_\textrm{LHS}^\star} \;
\mathrm{CI}_{T}^\textrm{wd}(x^\star).
\end{equation}
\textcolor{blue}{Inspired by the Active Learning MacKay criterion
\citep{seo2000alm}, we select the $x^\star$ with the largest uncertainty 
to drive exploration of $T(x^\star)$ across $\mathcal{X}^\star$. This, in turn, 
reduces uncertainty and increases the accuracy of our estimation over the 
course of successive acquisitions.} 

Then, we select 
$\mathbf{x}_{n+1}^{-\star}$ by maximizing PEI along the chosen slice:
\begin{equation}\label{eq:x-starnext}
\mathbf{x}_{n+1}^{-\star} = 
\argmax_{\mathbf{x}^{-\star}\in\{\mathbf{X}_\textrm{tri}\mid\mathbf{X}_n^{-\star}\}}\;
\mathrm{PEI}\left(\mathbf{x} = \left[x_{n+1}^\star, \mathbf{x}^{-\star}\right]\mid \mathcal{D}_n\right).
\end{equation}
This calculation is \textcolor{blue}{low-cost} since our tricands set is not 
too large and $\mu_{T}(x^\star_{n+1})$ has already been estimated. We then 
observe $y_{n+1} = f(\mathbf{x}_{n+1})$, update our surrogate, and repeat until 
our evaluation budget has been spent. Our complete PBO procedure is outlined in 
Algorithm~\ref{alg:profilebo}. Notice, although we use a 1D LHS through the 
acquisition loop, we prefer to use an evenly spaced grid in our final 
estimation.
\medskip

\begin{algorithm}[H] 		
	\DontPrintSemicolon
	{\bf Inputs:} Black-box simulator $f$, initial training data $\{\mathbf{X}_n, \mathbf{y}_n\}$, 
		trained surrogate $f_n$, maximum number of simulator evaluations $m$ \;

	\For{$i = (n+1), \dots, m$}{
		Generate $\mathbf{X}_{\textrm{tri}^+} = \mathbf{x}_\textrm{LHS}^\star
			\textcolor{blue}{\times}\left\{\mathbf{X}_{\textrm{tri}}\mid \mathbf{X}^{-\star}_n\right\}$ \;
		Estimate $\hat{T}(x^\star)$ \;
		Select $x_{n+1}^\star$ with largest uncertainty (Eq.~\ref{eq:xstarnext})\;
		Select $\mathbf{x}_{n+1}^{-\star}$ by maximizing PEI (Eq.~\ref{eq:x-starnext})\;
		Update $\mathbf{X}_n = \{\mathbf{X}_n, \mathbf{x}_{n+1}\}$ and 
			$\mathbf{y}_n = \{\mathbf{y}_n, f(\mathbf{x}_{n+1})\}$\;
		Retrain surrogate \;
		Increment $n = n+1$\;
	}
	{\bf Output:} Final $\hat{T}(x^\star)$ using $\mathbf{X}_{\textrm{tri}^+} = 
		\mathbf{x}_\textrm{grid}^\star\textcolor{blue}{\times}\left\{\mathbf{X}_{\textrm{tri}}\mid \mathbf{X}^{-\star}_n\right\}$ 
	\caption{Profile Bayesian Optimization.}
	\label{alg:profilebo}
\end{algorithm}

\medskip

\begin{figure}[H]
	\centering
	\includegraphics[width=.3\linewidth]{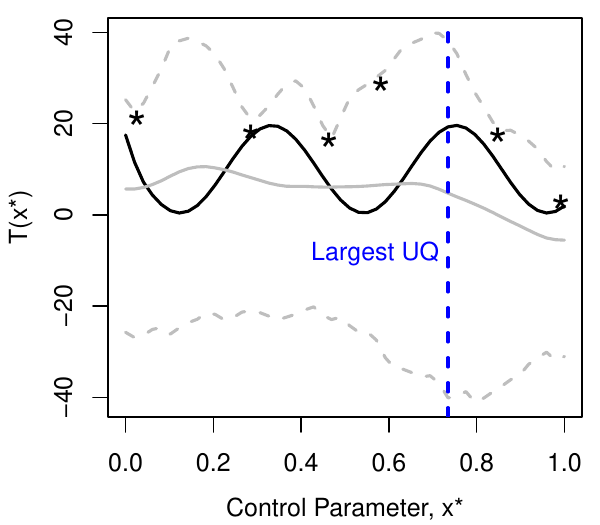}
	\includegraphics[width=.3\linewidth]{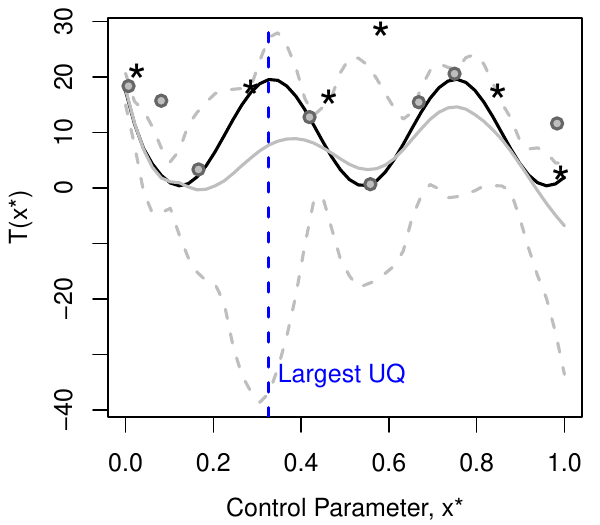}
	\includegraphics[width=.3\linewidth]{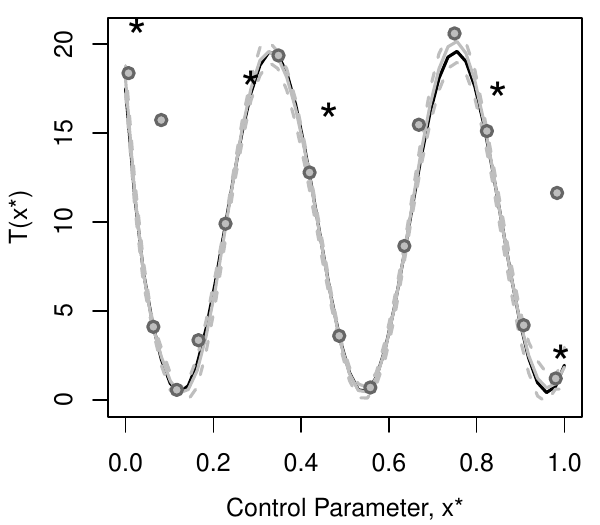}
	\caption{$\hat{T}(x^\star)$ (solid/dashed gray) for the Branin function 
		from an initial 10-point LHS (left), followed by 10 PBO acquisitions 
		(center), and another 10 PBO acquisitions (right). Black lines mark the 
		true $T(x^\star)$. Blue dotted lines mark the $x^\star$ values with the 
		widest uncertainty. Black stars and gray circles represent observations 
		from the initial LHS and PBO acquisitions, respectively.}
	\label{fig:pbo}
\end{figure}

Figure~\ref{fig:pbo} offers an inside look at this process for the Branin 
function.  Starting with $\hat{T}(x^\star)$ estimated from our 10-point LHS 
(left panel here, also shown earlier in Figure~\ref{fig:braninpost}), we select 
the $x^\star$ with the widest $\mathrm{CI}_T(x^\star)$, marked here by the blue 
dotted line. Then we choose the tricands point in the nuisance parameter space 
with the largest PEI on that slice.  The center and right panels show progress 
after 10 and 20 PBO acquisitions, respectively.  As the design progresses, our 
estimation of $T(x^\star)$ exhibits increased accuracy and decreased 
uncertainty.  The gray circles indicate data points that were selected by our 
PBO procedure (some of them are cut-off above, having $y>20$).  Many of the 
acquired points fall near the true minimum of their respective slice, 
indicating that PEI is effectively exploiting the minimum on the selected 
slices.

\section{Synthetic Examples}\label{sec:toyproblems}

In this section, we validate the performance of our proposed PBO against 
\textcolor{blue}{various} alternatives on a variety of benchmark exercises. 
We measure performance of the estimation of $T(x^\star)$ in four ways. 
First, to compare accuracy we assess average performance with root mean 
squared prediction error (RMSE; lower is better). Then, to understand 
the quality of our uncertainty quantification, we consider both the 
coverage and the average width of the 95\% credible intervals (the 
latter is denoted AvgCI). Narrower CIs are preferred as long as the 
coverage is high, preferably near 95\%. \textcolor{blue}{Worst-case 
performance is measured using maximum absolute deviation (MaxAD; lower is better);
results for this metric are included in Supplement~\ref{supp:results}.} 
Formal definitions of all metrics are provided in Supplement~\ref{supp:functions}.

In the following sections, we will consider stationary and nonstationary 
examples, demonstrating applicability to both GP and DGP surrogates. 
\textcolor{blue}{We restrict our test functions to those with unique profile
optima for each $x^\star$, with further discussion provided in 
Section \ref{sec:summary}.} For all surrogates, we 
use a fixed nugget of $1.0\times 10^{-6}$ to reflect the deterministic nature 
of the functions. Training budgets are adjusted based on the dimension and 
complexity of each test function. \textcolor{blue}{For stationary functions, we 
start with LHS designs of size $n=5d$; for more complex nonstationary 
functions, we start with LHS designs of size $n=10d$. For both stationary and 
nonstationary functions, we proceed until $m=3n$, although
we find convergence of the AvgCI often occurs before that point. While 
progress in maximum EI is commonly used in Bayesian optimization 
to measure convergence, we find AvgCI is better than PEI for assessing convergence
of our PBO procedure since we do not globally maximize PEI.} We prescale inputs to 
$\mathcal{X} = [0,1]^d$ and responses to have zero mean and unit variance. All 
exercises are repeated with 30 re-randomized starting designs. Reproducible 
code for all synthetic experiments is available in our public 
repository\footnote{\url{https://bitbucket.org/boothlab/rdc/}}. 
\textcolor{blue}{Computation times for each example are included in Supplement 
\ref{supp:time}.}

\subsection{Stationary Functions}\label{ss:stationary}

For our stationary examples, we fit GP surrogates with separable lengthscales 
and \textcolor{blue}{Mat\'{e}rn} kernels using the ``scaled Vecchia'' 
implementation of \citet{katzfuss2022scaled}, which is built upon the {\tt 
GpGp} \citep{GpGp} and {\tt GPvecchia} \citep{GPvecchia} R packages. We do not 
need Vecchia approximation for training purposes given the small datasets, but 
we do use Vecchia approximation for fast joint posterior sampling. Throughout, 
we use a conditioning set size of 40 when drawing Vecchia-approximated samples.

We compare our PBO procedure to a random LHS of equivalent size, a typical 
Bayesian optimization using multi-start gradient-based maximization of expected 
improvement, and the PEI approach of \citet{ginsbourger2013profile}. The LHS 
serves as a baseline only---its design is not sequential, and it does not use 
surrogates to select design locations. The typical BO serves as a different 
benchmark, displaying the consequences of targeting global optima over profile 
optima. \textcolor{blue}{PEI, as originally proposed in 
\citet{ginsbourger2013profile}, relies on grids as candidates and is not 
feasible above 2 dimensions. To enable the use of PEI as a comparator in higher 
dimensions, we use our modified tricands as potential PEI acquisitions.}

\textcolor{blue}{For all methods, we estimate $T(x^\star)$ as outlined in 
Section \ref{ss:gp} by obtaining 1,000 posterior samples at our modified tricands 
locations.  Because all procedures borrow our modified tricands in some fashion, 
comparisons of computation times are not enlightening.  We reserve further 
discussion for Supplement \ref{supp:results}.}

\subsubsection*{2D Branin}

First, we revisit the 2D Branin function shown earlier in 
Figure~\ref{fig:branin2D}, starting with a random LHS of size $n=10$ 
and ending with a total budget of $m=30$.  Figure~\ref{fig:pbo} showed 
the resulting $\hat{T}(x^\star)$ from one repetition of our PBO procedure. 
Similar figures for the LHS, BO, and PEI methods are provided in 
Supplement~\ref{supp:results}. 

\begin{figure}[H]
	\centering
	\includegraphics[width=0.375\linewidth]{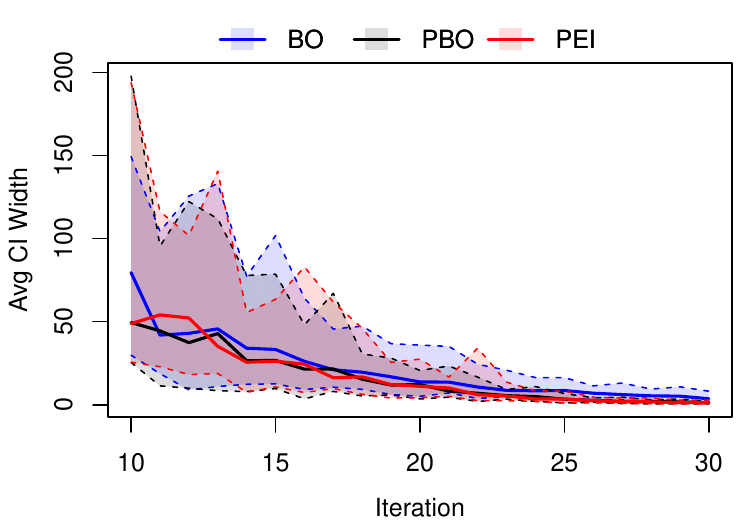}
	\includegraphics[width=0.375\linewidth]{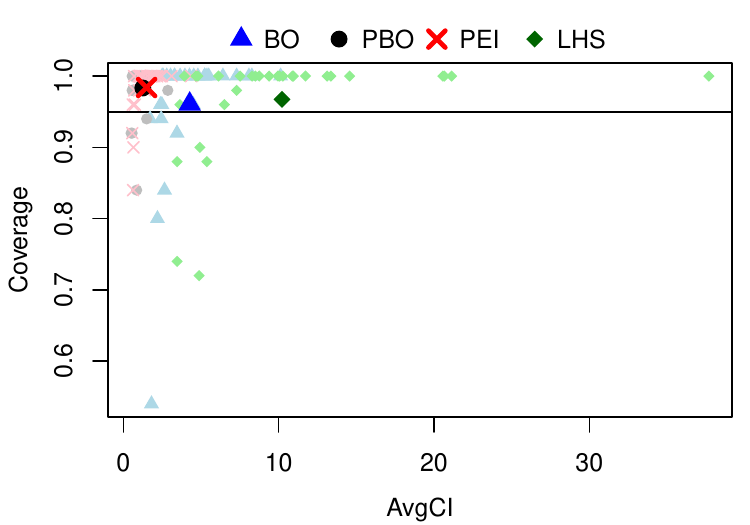}
	
	\includegraphics[width=0.375\linewidth]{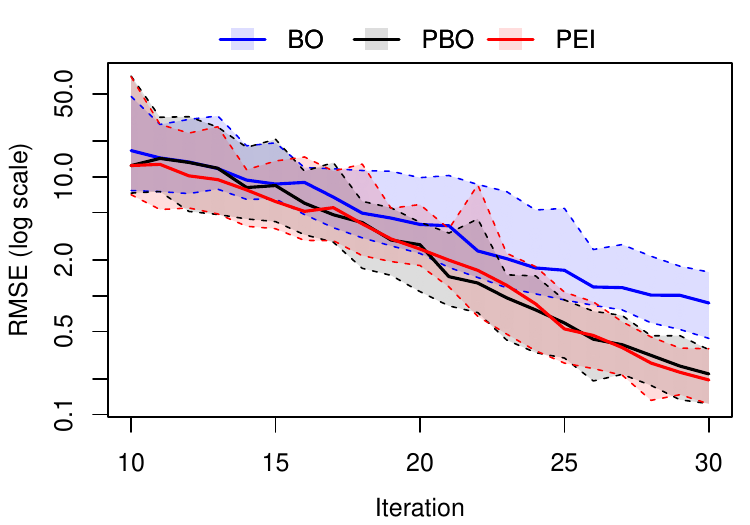}
	\includegraphics[width=0.375\linewidth]{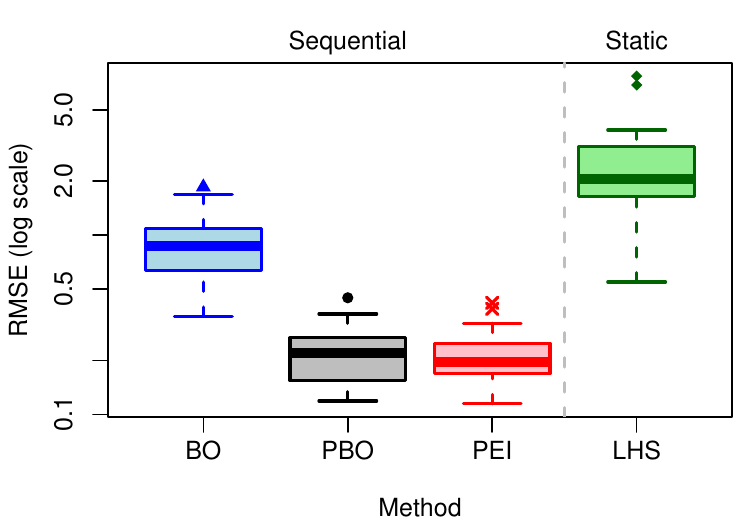}
	\caption{\textcolor{blue}{Results for the Branin function across 30 repetitions with $n=10$ and 
		$m=30$. Left two panels display average CI width and RMSE from $n$ to $m$
		with solid lines representing the mean and shaded intervals representing 95\% quantiles.
		Right two panels display static results at $m$. Large colored symbols in 
		the top right panel represent mean performance.}}
	\label{fig:comparisons2D}
\end{figure}

\textcolor{blue}{Figure~\ref{fig:comparisons2D} shows the performance of each
method across 30 re-randomized starting designs.  Each method offered
similar convergence in AvgCI (top left panel).  PBO and PEI achieved the
lowest RMSE (bottom panels) and the lowest AvgCI while maintaining over 
95\% coverage (top right panel). BO 
performs slightly worse as it is not designed 
to target the peaks in the Branin function's $T(x^\star)$. 
LHS offers the worst RMSE and the largest uncertainty, though it has
comparable coverage.  This is expected because the LHS designs do not 
attempt to target any minima.}

\subsubsection*{3D Kyger}

\begin{figure}[H]
	\centering
	\includegraphics[width=0.375\linewidth]{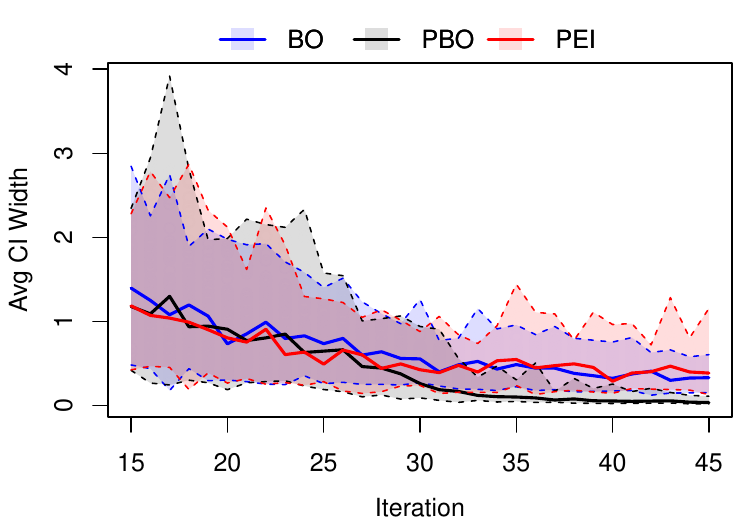}
	\includegraphics[width=0.375\linewidth]{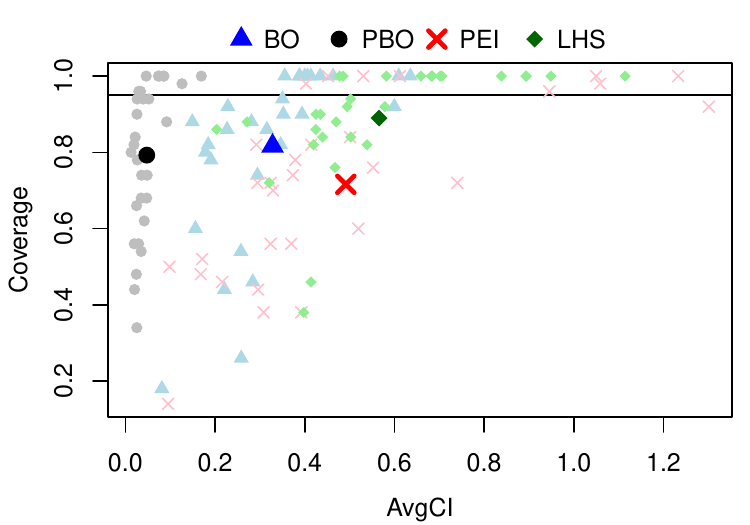}
	
	\includegraphics[width=0.375\linewidth]{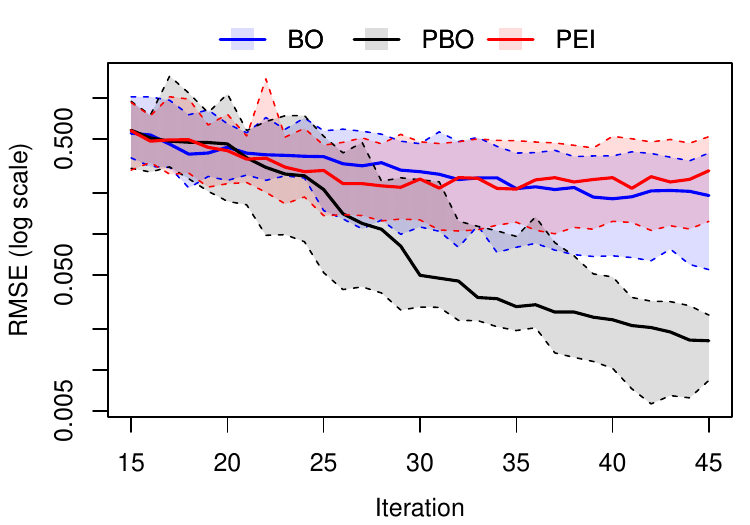}
	\includegraphics[width=0.375\linewidth]{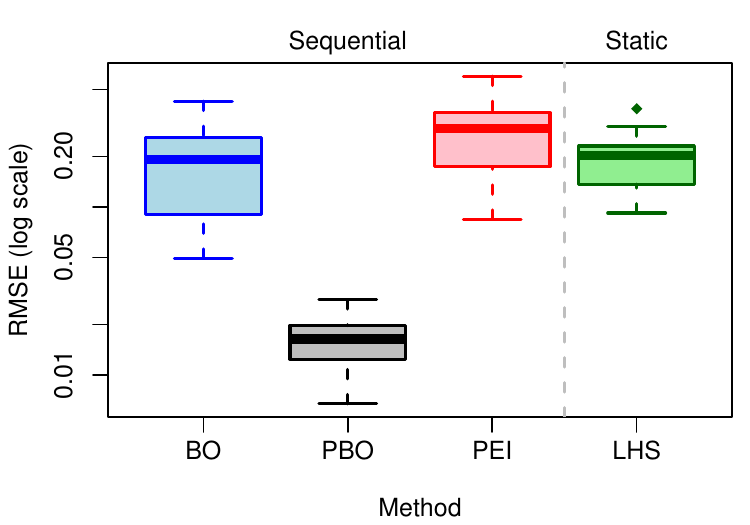}
	
	\caption{\textcolor{blue}{Results for the Kyger function across 30 repetitions 
	with $n=15$ and $m=45$, in the same format as Figure \ref{fig:comparisons2D}.}}
	\label{fig:comparisons3D}
\end{figure}

\begin{figure}[H]
	\centering
	\includegraphics[width=.375\linewidth]{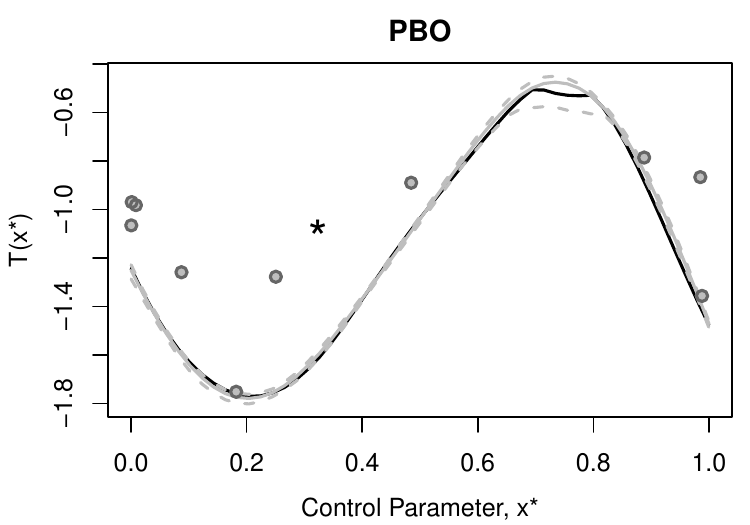}
	\includegraphics[width=.375\linewidth]{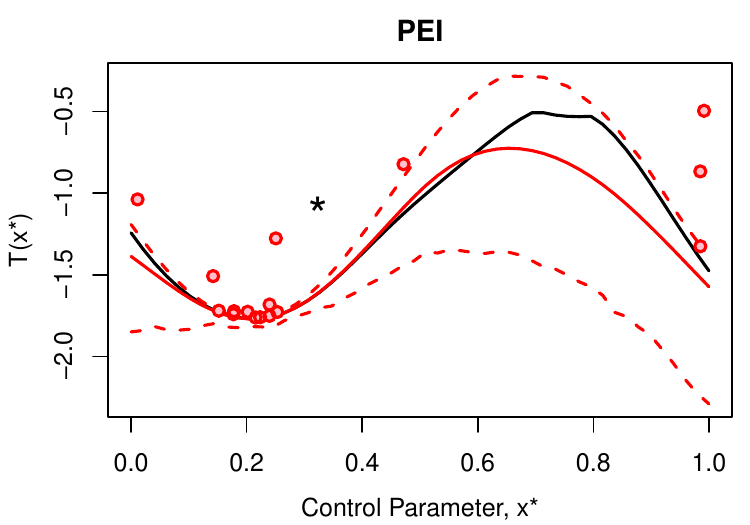}
	\caption{\textcolor{blue}{PBO and PEI estimation of $T(x^\star)$ at $m=45$
	for the 3D Kyger function. Black stars indicate initial LHS points, and 
	colored circles indicate acquired points. 
	Solid black lines denotes the true $T(x^\star)$, solid colored lines 
	denote $\mu_{T}(x^\star)$, and dashed colored lines denote 
	$\mathrm{CI}_{T}(x^\star)$.}}
	\label{fig:Tstar3D}
\end{figure}

Our second stationary example is the 3D ``Kyger'' function, defined in 
Supplement~\ref{supp:functions}. We begin with $n=15$ and end with $m=45$.  
Results are shown in Figure~\ref{fig:comparisons3D}. \textcolor{blue}{The top 
left panel, which displays the convergence of AvgCI, now shows convergence of 
PBO at widths much narrower than BO or PEI. While the coverage of PBO is 
slightly below target (top right panel), no method has ideal 
coverage regardless of their ending AvgCI. The lower two panels show 
PBO has measurably better convergence of RMSE and ending RMSE compared to the 
remaining methods. Figure~\ref{fig:Tstar3D} shows the estimated $T(x^\star)$
for a single run of PBO and PEI.  Due to the exploitative nature of PEI---similar to its 
predecessor, EI---it overfocuses on the global minimum, spending over half of 
its acquisition budget in this region. This leads to a lack of exploration and poor 
estimation of the remainder of the $T(x^\star)$.} PBO performs better because 
its two-stage ``explore then exploit'' acquisition procedure strikes an 
effective balance.

\subsection{Nonstationary Function} \label{sec:nonstationary}

For our nonstationary example, we use the {\tt deepgp} R package \citep{deepgp}
to fit two-layer DGPs with \textcolor{blue}{Mat\'{e}rn} kernels. We conduct 
10,000 MCMC iterations for initial fits and 2,000 for subsequent fits after 
each new acquisition (initializing $\mathbf{W}_n^{(0)}$ and all hyperparameters 
with the last sampled values from the previous fit). We also set 
$\boldsymbol\mu_w = \textbf{X}$ (Eq.~\ref{eq:dgp}). We remove half of the MCMC 
iterations for burn-in, then thin down to 100 iterations. Joint posterior 
samples are generated from the modified tricands set with 10 samples coming 
from each MCMC iteration, for a total of 1,000. To provide insight into the 
benefit of the DGP's flexibility, we compare our PBO procedure with a 
stationary GP against the same procedure with the nonstationary DGP.  We drop 
BO and PEI as competitors given their poor performance in 
Section~\ref{ss:stationary}, but we retain the LHS benchmark, using both GP and 
DGP surrogates on the same LHS designs.

\subsubsection*{4D Squiggle}

\begin{figure}[H]
	\centering
	\includegraphics[width=0.375\linewidth]{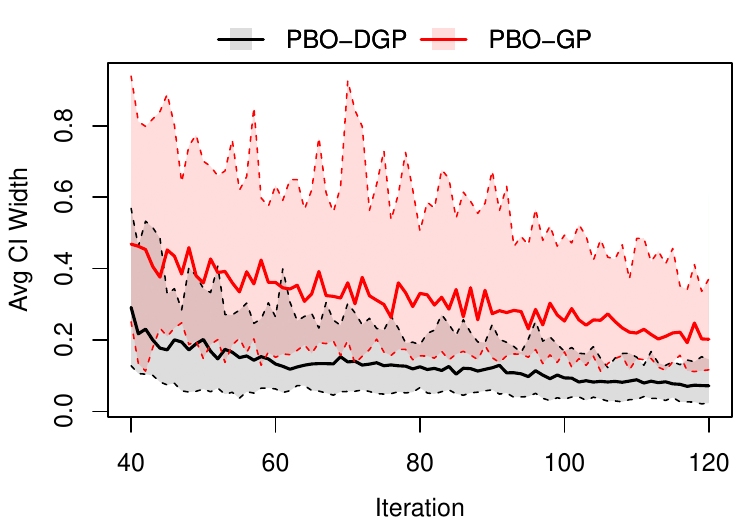}
	\includegraphics[width=0.375\linewidth]{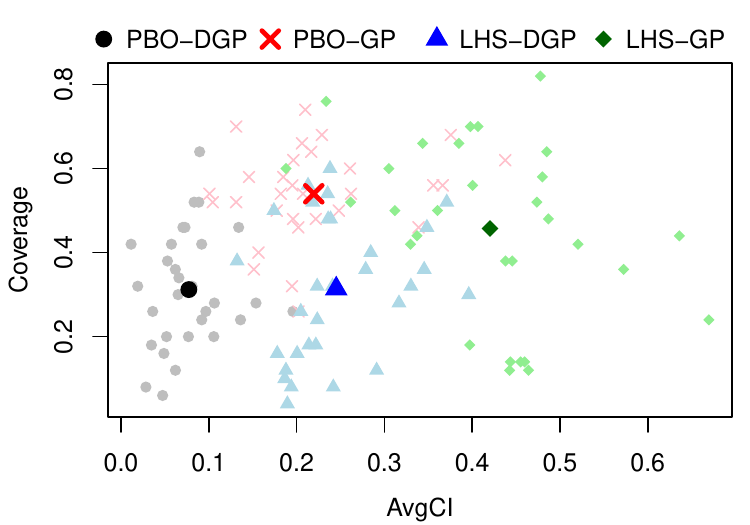}
	
	\includegraphics[width=0.375\linewidth]{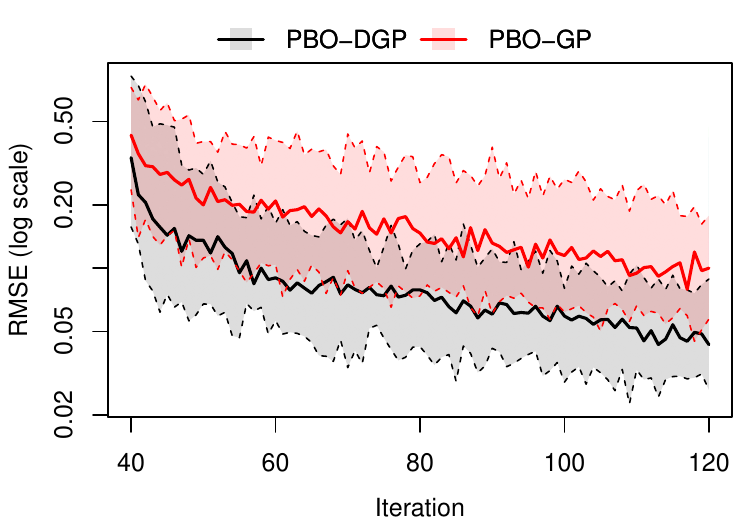}
	\includegraphics[width=0.375\linewidth]{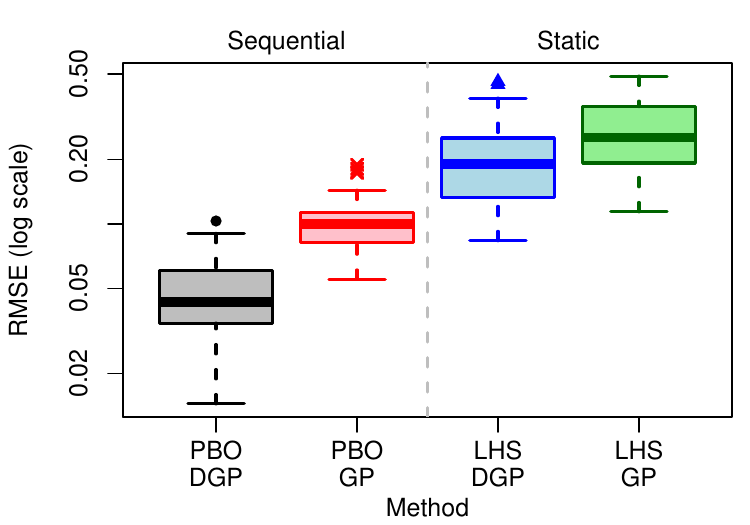}
	\caption{\textcolor{blue}{Results for the nonstationary squiggle function 
		across 30 repetitions with $n=40$ and $m=120$. Left two panels display 
		average CI width and RMSE from $n$ to $m$; right two panels display 
		static results at $m$.}}
	\label{fig:comparisons4DN}
\end{figure}

We use the nonstationary ``squiggle'' function \citep{duqling, 
rumsey2025all}, which we have adapted from its original 2D form to create a 
similar 4D function (Supplement~\ref{supp:functions}). We begin with $n=40$ and 
end with \textcolor{blue}{$m=120$}, with results presented in 
Figure~\ref{fig:comparisons4DN}. \textcolor{blue}{The LHS DGP outperforms the
LHS GP given the nonstationary nature of the function, but both are beaten by
strategic PBO designs.  The PBO DGP outperforms the PBO GP, offering consistently
lower RMSE.  The PBO DGP shows much tighter uncertainty bounds across the design
(top left), but this results in lower coverage (top right).  None of the comparators
reached satisfactory coverage, which we attribute to the complicated nature
of the response surface.  The overconfidence 
of the DGP predictions are of note, though the 
performance of the RMSE and MaxAD (in Supplement~\ref{supp:results}) show
the predictions are still more accurate.}

\section{RDC-Diffuser}\label{sec:diffuser}

\begin{figure}[H]
	\centering
	\includegraphics[width=0.8\linewidth]{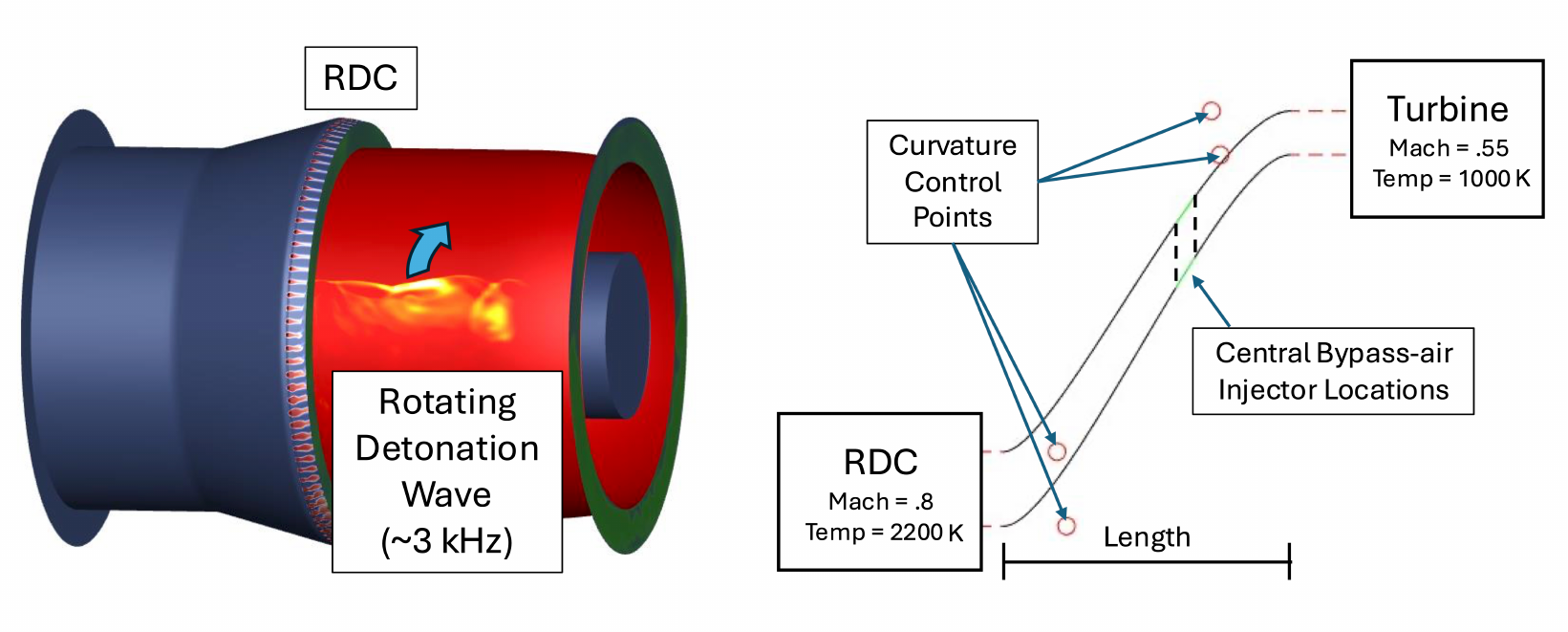}
	\caption{{\it Left:} Schematic of a detonation wave traveling through an 
		RDC. {\it Right:} Parameterized geometry for a 2D slice of the 
		diffuser. Flow speed and temperature are reduced upon exiting the 
		diffuser.}
	\label{fig:diffuser}
\end{figure}

Rotating detonation combustors utilize a circumferentially rotating 
detonation wave to consume fuel and initiate combustion, as depicted in the 
left panel of Figure \ref{fig:diffuser}. RDCs may increase thermodynamic 
efficiency up to 20\% when compared to standard conventional combustors, which 
could reduce fuel consumption within an engine by up to 9\% with the same power 
output \citep{jones2013potential}. Power extraction requires the flow exiting a 
combustor to enter a gas turbine, but this process poses several challenges. 
Detonation combustion invokes high temperatures, presenting issues with thermal 
management \citep{heiser2002thermodynamic}. The rotating detonation wave 
travels along the walls of the combustor at high speeds, resulting in high 
frequency pulses and permeations within the flow which can inhibit power 
extraction \citep{fernelius2013effect}. Also, the flow speed must be reduced to 
appropriate levels for the provided turbine. 

The diffuser, which will mix the hot exhaust and cold bypass-air downstream of 
the RDC, is designed to condition and cool the flow and reduce negative 
oscillatory effects from the rotating detonation wave. Diffusers hold great 
potential to improve the efficiency of power extraction from RDCs. 
\citet{naples2014rotating} replaced the stock combustor of a T63 helicopter 
engine with an RDC-diffuser and found that oscillations were damped by 60-70\% 
and hot exhaust temperatures were reduced to near ideal levels. 
\citet{grunenwald2023investigation} explored the integration of an RDC-diffuser 
into a Rolls Royce M250 engine and found that non-axial flows could be forced 
into a more linear regime through a ``swan-neck'' shaped diffuser, resulting in 
a more uniform flow.

\begin{figure}[H]
	\centering
	\includegraphics[width=0.9\linewidth]{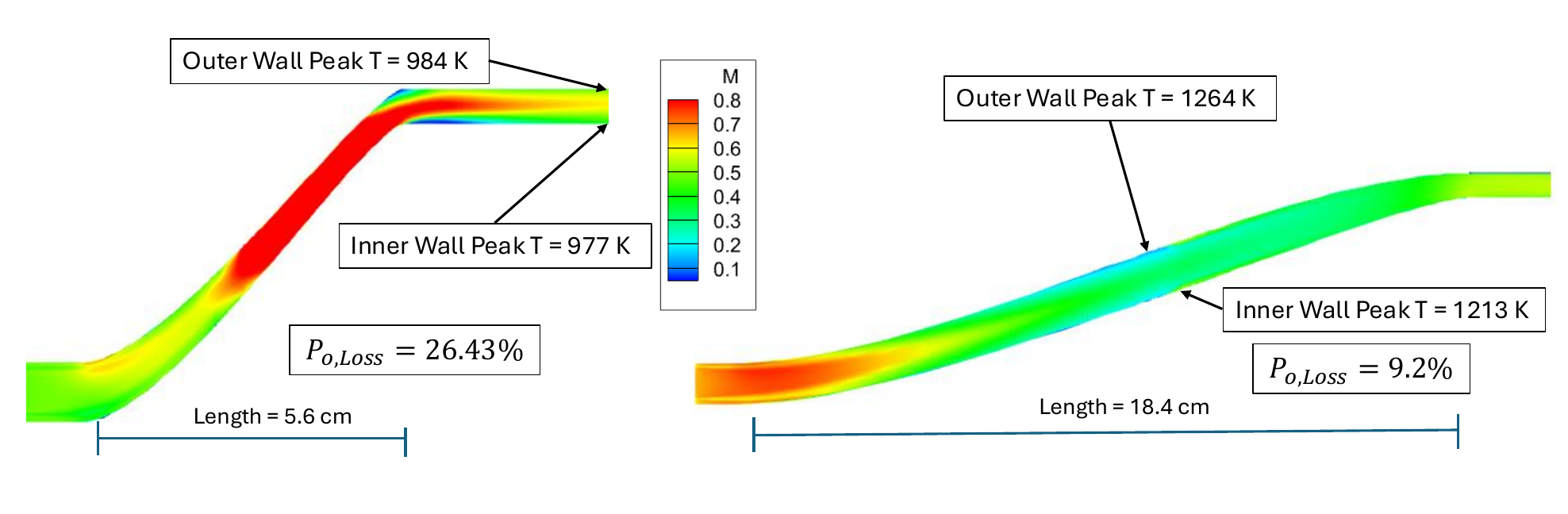}
	\caption{Pressure loss (P) with Mach number (M) and temperature (T) for
		two runs of the 6D diffuser simulation with a short length (left)
		and a long length (right).}
	\label{fig:diffuser2}
\end{figure}

While diffusers show great promise, the potential for them to improve RDC power 
efficiency is contingent on effective design. Here, we work with a computer 
simulation that returns the pressure lost through diffusion as a function of 6 
design parameters: length, location of cold bypass-air inlets, and 4 control 
points that determine the curvature. \textcolor{blue}{Pressure loss is 
equivalent to energy loss.} The right panel of Figure \ref{fig:diffuser} offers 
a simple representation of the diffuser and its parameterization.  To 
demonstrate the significance of the diffuser's length on energy efficiency, 
Figure \ref{fig:diffuser2} depicts two 2D diffusers of varying lengths, but 
with identical normalized curvatures. The short diffuser has a large pressure 
loss of 26.43\%, indicating poor energy efficiency. The longer diffuser has a 
low pressure loss of 9.2\%, but the temperatures of the flow at the walls 
exceeds 1000K, \textcolor{blue}{which could cause downstream parts to melt or 
warp.}

\begin{figure}[H]
	\centering
	\includegraphics[width=0.4\linewidth]{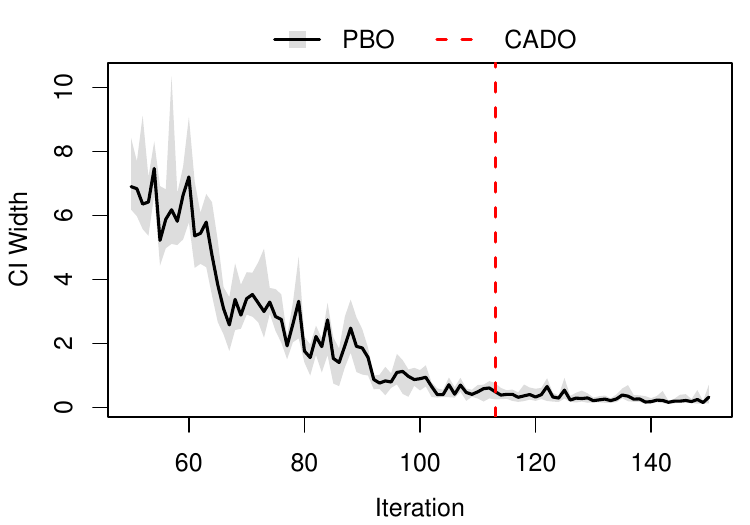}
	\includegraphics[width=0.4\linewidth]{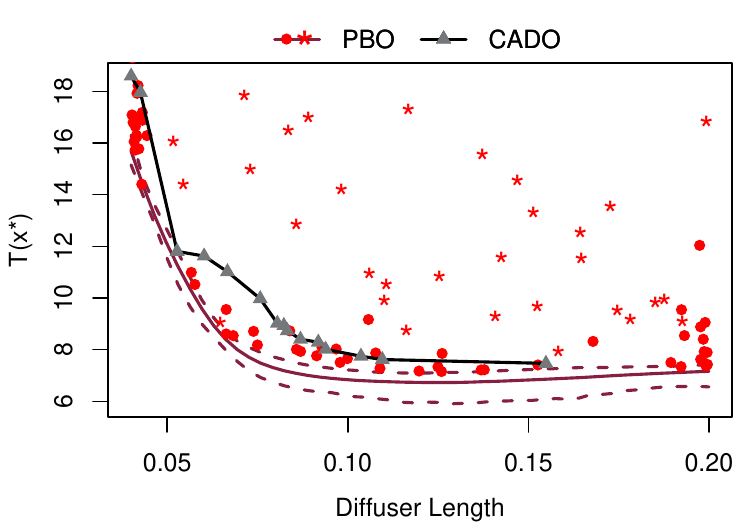}
	\caption{\textcolor{blue}{\textit{Left:} CI minimum/average/maximum width
	across acquisitions for our DGP PBO procedure on the RDC simulation.  Vertical
	red line marks the number of simulations used by the comparable evolutionary
	optimzer (CADO).  \textit{Right:} Estimated profile optima $\hat{T}(x^\star)$ 
	(solid/dashed maroon) after a 50-point LHS and 63 acquisitions, the same budget 
	used by CADO. Red stars indicate initial LHS points; circles indicate PBO 
	acquisitions.  Black line and gray triangles indicate the Pareto front
	found by CADO.}}
	\label{fig:diffuserresults}
\end{figure}

We implement our PBO procedure to identify the profile optima in pressure loss 
as a function of the diffuser's length (leaving 5 nuisance parameters).  The 
complex dynamics of the RDC-diffuser power output warrant the nonstationarity 
of a DGP surrogate. \textcolor{blue}{We begin with a 50-point LHS, indicated by 
the red stars in the right panel of Figure~\ref{fig:diffuserresults}. Only the 
lowest pressure loss points are visible; designs with larger pressure losses 
are cut off above. Then, we fit a two-layer DGP as described at the beginning of 
Section~\ref{sec:nonstationary}, and estimate the resulting $T(x^\star)$. 
The minimum, maximum, and average CI width for our estimate is displayed 
in the left panel of Figure~\ref{fig:diffuserresults} (due to the computational
expense of the simulation, we only have a single run).  The UQ is large 
for this initial estimation due to the input dimension ($d = 6$) and 
limited sample size ($n=50$).}

\textcolor{blue}{We proceed with 100 acquisitions until we reach
$m = 150$ (following the $m=3n$ rule used in Section~\ref{sec:toyproblems}).
The left panel of Figure~\ref{fig:diffuserresults} shows progress in the CI
widths across acquisitions.  While we do not know the truth, the convergence 
observed after 100 acquisitions is a good indication that our procedure is 
providing reliable estimates. Fortunately, we can compare our 
results to the multi-objective optimization procedures which were standard
practice in this application area before this work.  We use an
evolutionary computer-aided design optimizer (CADO) that seeks to jointly
minimize the pressure loss and diffuser length \citep{braun2021aerothermal,verstraete2010design}.
This optimization terminated after 113 simulator evaluations.  To provide an apples-to-apples
comparison, the right panel of Figure~\ref{fig:diffuserresults} shows the estimated
$T(x^\star)$ from our PBO procedure after 113 total acquisitions, with the Pareto
front found by the multi-objective optimization overlaid.
Our PBO procedure provides estimation and UQ across the entire range of diffuser
lengths and found many designs with pressure losses lower than the Pareto front 
found by CADO, indicating promising designs for practical high efficiency engines.
The results from our PBO procedure after the complete $m=150$ evaluations 
are provided in Supplement~\ref{supp:results}.}

\section{Conclusion}\label{sec:summary}

Our aim was to identify the profile optima (Eq.~\ref{eq:definetx}) of a 
deterministic black-box computer simulation over the full support of a single 
control parameter ($x^\star\in\mathcal{X}^\star$) given a limited evaluation 
budget. Traditional Bayesian optimization is not designed to explore the full 
control parameter space, making it ill-equipped for profile optimization. Prior 
work on profile optimization was limited to two-dimensional functions and 
similarly tended to over-exploit near global optima. We proposed a novel 
two-stage surrogate-informed acquisition method, termed profile Bayesian 
optimization. The first step forces exploration of $\mathcal{X}^\star$ by 
identifying the $x^\star$ with the largest uncertainty in the surrogate's 
estimation of $T(x^\star)$. The second step selects nuisance parameter values 
by maximizing profile expected improvement along the chosen slice.  Our 
two-stage approach is feasible in higher dimensions and is compatible with 
nonstationary DGP surrogates, which can provide superior predictions of complex 
functions. Our procedure performed favorably on various synthetic examples, 
improving upon space-filling designs as well as existing BO and profile 
optimization methods.  On our motivating computer experiment, our procedure 
provided a useful estimate of the profile optima from only 150 simulator 
evaluations.

There are several interesting avenues for future research in this area. 
\textcolor{blue}{Our methodology is designed for functions that have a unique 
profile optimum for each $x^\star$ (i.e., each diffuser length, or $x^\star$ 
``slice'', should have one optimal design, or global optimum in 
$\mathcal{X}^{-\star}$). For functions with non-unique global optima for any 
value of $x^\star$, our methodology is designed to identify the minimum 
attainable response value but not necessarily explore all $\mathbf{x}^{-\star}$ 
values that produce that same minimum for a particular $x^\star$. Ensuring 
exploration of multiple global optima for a particular $x^\star$ will likely 
require adaptations to our current methodology; we leave this as an avenue for 
future work.}

There is also potential to improve our modified tricands procedure, particularly 
in the allocation of fringe candidates. In our \textcolor{blue}{examples}, 
we set fringe candidates to be placed at 90\% of the distance to the boundary. 
This gets close to corners and borders, but it avoids placing candidates 
squarely on the boundary. If the profile optima is found in a corner, our 
candidates could struggle to find them quickly.  Additionally, as input 
dimension increases (beyond what we considered here), tricands may become 
computationally infeasible and suboptimal in their coverage. In these cases, 
adaptations like Voronoi candidates \citep{wycoff2025voronoi} may be 
necessary. \textcolor{blue}{Moreover, it is worth investigating the selection
of appropriate starting design sizes and acquisition budgets.  We followed 
standard procedures in our exercises ($n=5d$ or $n=10d$ with $m=3n$), but
convergence was often achieved before we reached the end of our designs.}

\textcolor{blue}{In our nonstationary test function, we noticed an interesting 
phenomenon: our more flexible DGP surrogates, although offering the best 
accuracy, suffered from low coverage.  We suspect that increasing the number 
of posterior samples used to estimate $T(x^\star)$ could improve the coverage,
but the runtime required to generate these samples increases exponentially and 
becomes nearly infeasible.  We provide further evidence and discussion
in Supplement~\ref{supp:time}.  Alternative solutions require further
investigation.}

There are many avenues for extensions in our motivating application. To improve 
surrogates of our RDC-diffuser simulation, we will explore 
gradient-enhancement, which can offer huge improvements in surrogate 
performance from the same budget of simulator evaluations.  This will involve 
upgrading our simulation to return gradients through the use of adjoint 
solvers, then employing gradient-enhanced DGP surrogates 
\citep{booth2025deep}.  Extensions of our PBO procedure that incorporate 
temperature constraints would be relevant, as excessive temperatures put the 
diffuser in danger of melting.  We also plan to leverage Bayesian calibration 
\citep{kennedy2001bayesian} to calibrate our computer simulation with data from 
physical lab experiments, where noise and bias may be present. Ultimately, our 
RDC-diffuser simulation has potential to serve as a digital twin 
\citep{willcox2023foundational} of a real RDC-diffuser in an engine.

\subsection*{Acknowledgements}

This work was supported by the U.S. National Science Foundation under 
Award Number 2533443.

\bibliographystyle{jasa}
\bibliography{main}

\newpage
\begin{center}
{\Large\bf SUPPLEMENTARY MATERIAL}

\end{center}
\appendix

\section{Test Functions}\label{supp:functions}

In this section we provide the formulaic details for the test functions
used in Section \ref{sec:toyproblems}.  For all functions, we prescale
inputs to $[0,1]^d$ and responses to zero mean with unit variance
before modeling.
\newline\newline
\noindent The Branin function \citep{surjanovic2013virtual} is defined as
\begin{equation*}
	f(\textbf{x}) = a\left(x_2-bx_1^2+c_x1-r\right)^2 + s\left(1-t\right)\cos\left(x_1\right) + s
	\quad\textrm{for}\quad
	x_1\in[-5,10]
	\quad
	x_2\in[0,15].
\end{equation*}
The 3D Kyger function is defined as:
\begin{equation*}
	\begin{aligned}
	f(\textbf{x}) &= \exp\left(-x_1-\cos(2\pi x_1)\right)+
		\sin\left(2\pi x_3\right)+
		\exp\left(-x_3*\sin(2\pi x_1)\right)+\\
	&\quad\quad \cos\left(2\pi x_2\right) - \exp\left(-x_2*\cos(2\pi x_1)\right)
	\quad\textrm{for}\quad
	x_1,x_2,x_3\in[0, 1].
	\end{aligned}
\end{equation*}
The squiggle function \citep{duqling,rumsey2025all} is defined as:
\begin{equation*}
		f(\textbf{x}) = x_1\phi\left(\frac{\sum_{i=2}^4 x_i^2 - \left[\sin\left(2\pi x_1^2\right)/4 - x_1/10 +0.5\right]}{\sigma}\right)\quad\textrm{for}\quad
		x_1,x_2,x_3,x_4\in[0.1,1],
		\quad
		\sigma=0.2,
\end{equation*}
where $\phi$ is the standard Gaussian probability density function.

\section{Performance Metrics}\label{supp:metrics}

Let $\mathbf{x}_\textrm{grid}^\star$ represent an evenly spaced grid of size 
$n_g$ covering the control parameter space $\mathcal{X}^\star$.  We obtain 
estimates of $T(x^\star)$ at each $x^\star\in\mathbf{x}_\textrm{grid}$ 
(Algorithm~\ref{alg:profilebo}).  Let $\mu_{T}(x^\star)$ denote the mean and 
$\mathrm{CI}_{T}(x^\star)$ denote the 95\% CI, with a width of 
$\mathrm{CI}_T^\textrm{wd}(x^\star)$.  Our performance metrics are defined as 
follows:

\begin{equation*}
\begin{aligned}
	\mathrm{RMSE} &= \sqrt{\frac{1}{n_g}\sum_{x^\star\in\mathbf{x}_\textrm{grid}^\star} 
		\left(\mu_{T}(x^\star)-T(x^\star)\right)^2} \\
	\mathrm{MaxAD} &= \max_{x^\star\in\mathbf{x}_\textrm{grid}^\star}
	\left(\left|\mu_{T}(x^\star) - T(x^\star)\right|\right) \\
	\mathrm{AvgCI} &= \frac{1}{n_g}\sum_{x^\star\in\mathbf{x}_\textrm{grid}^\star}
		\mathrm{CI}_{T}^\textrm{wd}(x^\star) \\
	\mathrm{Coverage} &= \frac{1}{n_g}\sum_{x^\star\in\mathbf{x}_\textrm{grid}^\star}
	\mathbbm{1}_{\left\{T(x^\star) \in \mathrm{CI}_{T}(x^\star)\right\}}
\end{aligned}
\end{equation*}

\newpage

\section{Additional Results}\label{supp:results}

\begin{figure}[H]
	\centering
	\includegraphics[width=.4\linewidth]{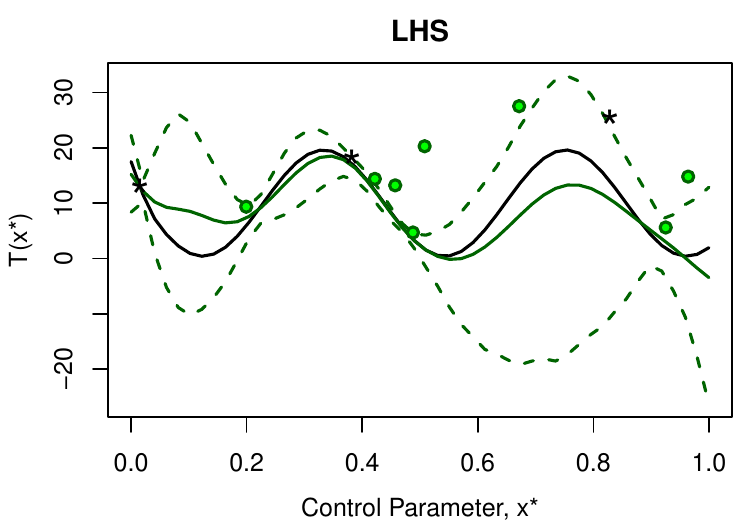}
	\includegraphics[width=.4\linewidth]{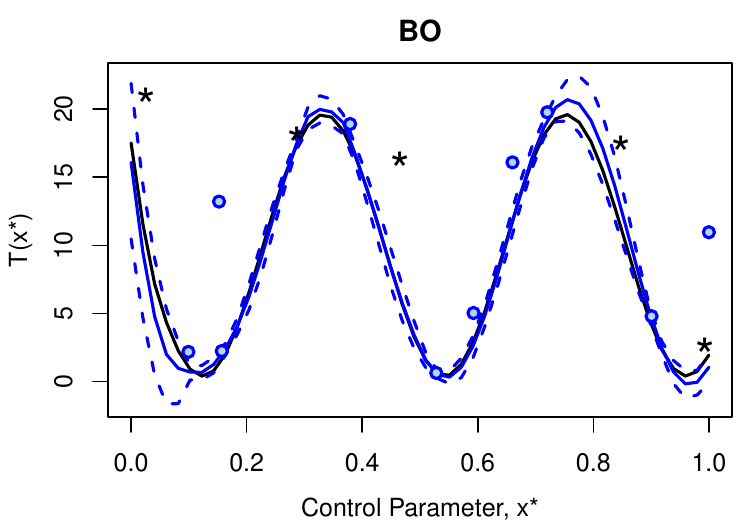} \\
	\includegraphics[width=.4\linewidth]{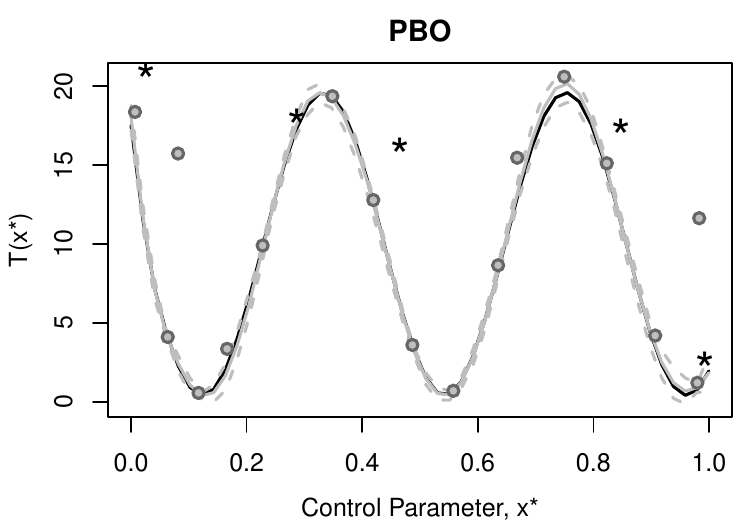}
	\includegraphics[width=.4\linewidth]{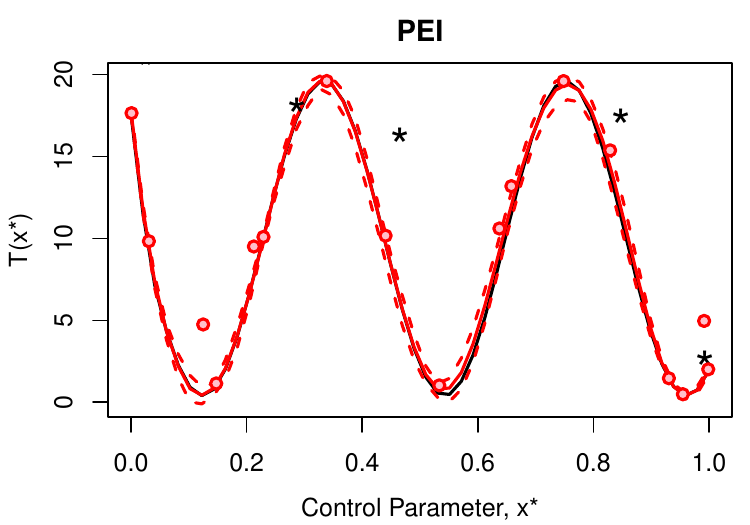}
	\caption{Estimation of $T(x^\star)$ for the Branin Function.  Truth
	in solid black.  Colored solid/dashed lines denote $\mu_{T}(x^\star)$ and
	$\mathrm{CI}_{T}(x^\star)$ for each method.  Black stars indicate initial LHS
	observations; colored circles indicate acquired points.}
	\label{fig:acqestbranin}
\end{figure}

\textcolor{blue}{First, we present the estimated $\hat{T}(x^\star)$ for the 2D
Branin function from each of the 
four methods mentioned in Section~\ref{sec:toyproblems}---the benchmark LHS, 
BO, PEI, and our method (PBO). Figure~\ref{fig:acqestbranin} displays these 
four estimations. All methods start with the same initial LHS of size $n=10$ 
and proceed with 20 acquisitions for a total of $m=30$; the benchmark LHS uses 
the static size $m$.}

\begin{figure}[H]
	\centering
	\includegraphics[width=0.375\linewidth]{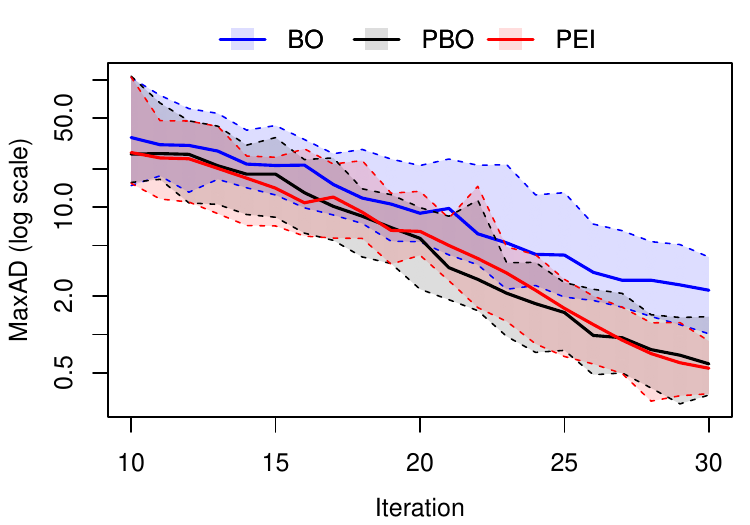}	
	\includegraphics[width=0.375\linewidth]{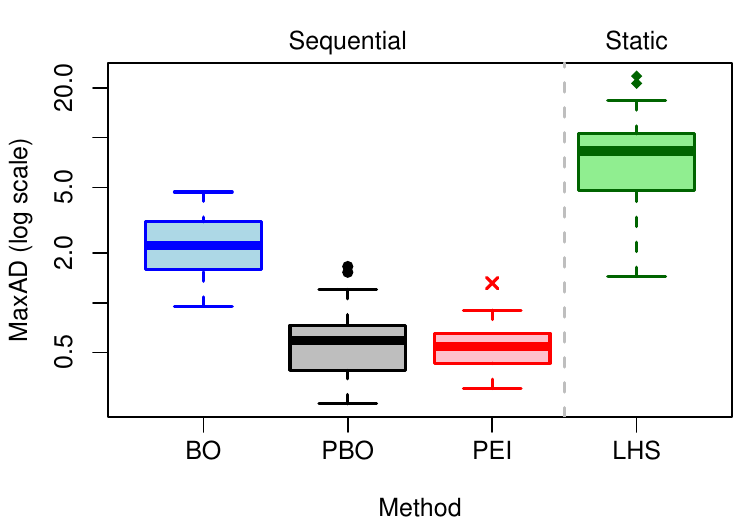}
	\includegraphics[width=0.375\linewidth]{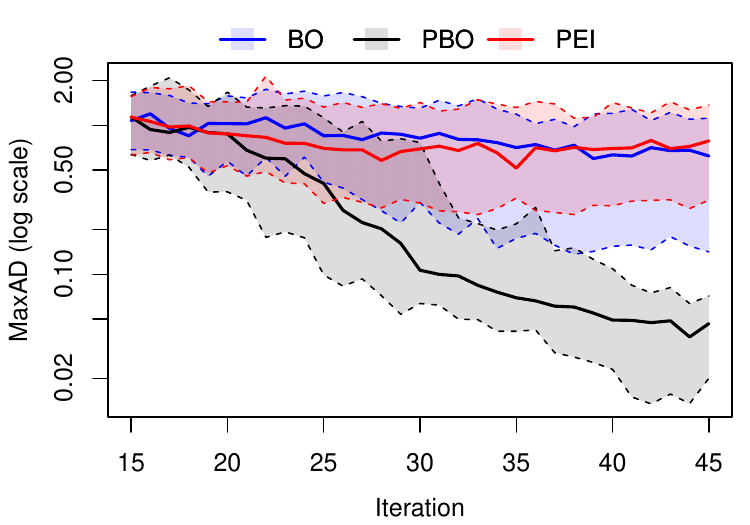}
	\includegraphics[width=0.375\linewidth]{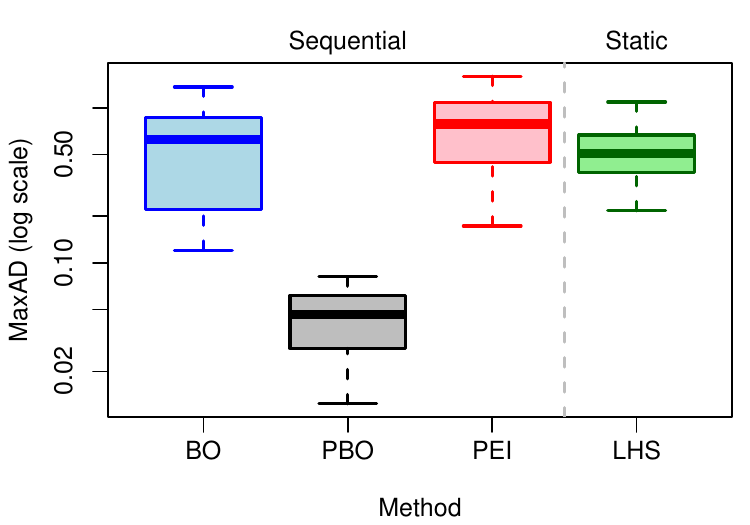}
	\includegraphics[width=0.375\linewidth]{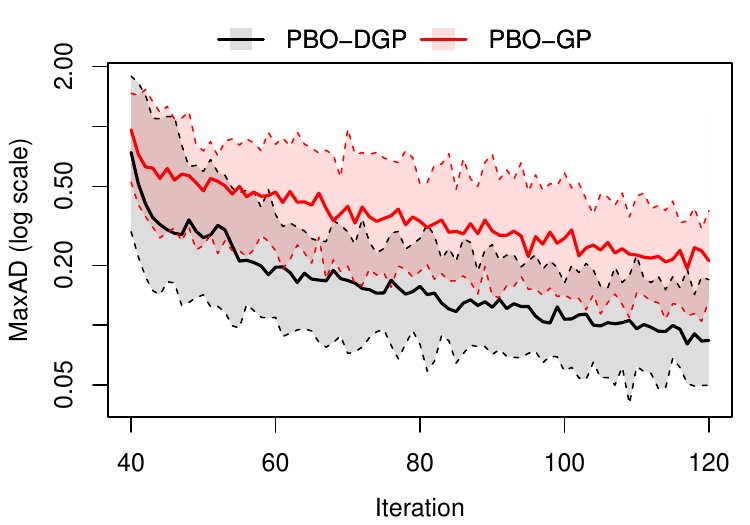}
	\includegraphics[width=0.375\linewidth]{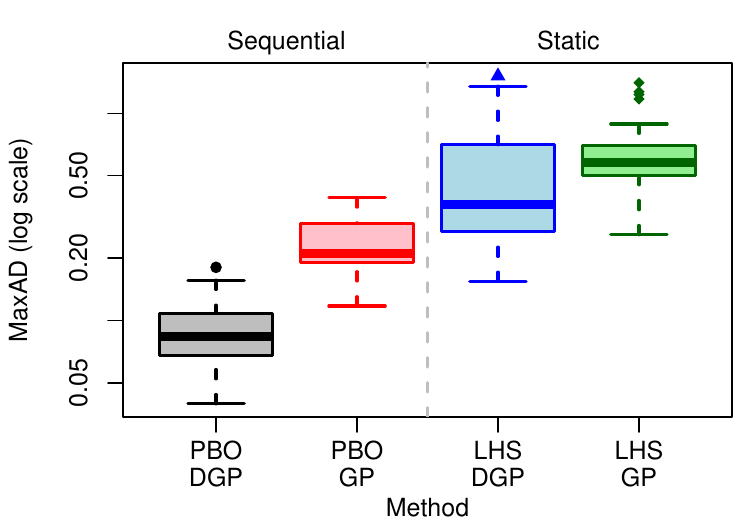}

	\caption{\textcolor{blue}{MaxAD progress over iterations (left panels) and 
	resulting MaxAD at final iteration (right panels) for the Branin function (top), 
	Kyger function (middle), and nonstationary squiggle function (bottom).  Simulation
	settings follow those described in Section~\ref{sec:toyproblems}.}}
	\label{fig:maxad}
\end{figure}

\textcolor{blue}{Second, we present the MaxAD performance for the three functions 
tested in Section~\ref{sec:toyproblems}.  Figure~\ref{fig:maxad} shows progress
from $n$ to $m$ (left panels) as well as the distribution of MaxAD at the
end of the design.  These results mirror the RMSE results reported in 
Section~\ref{sec:toyproblems}.  MaxAD for PBO and PEI is very similar for the Branin 
function (top row).  PBO has the best performance on the 3D Kyger 
function (middle row).  The DGP PBO method outperforms the GP PBO and static LHS 
benchmarks on the nonstationary squiggle function (bottom row).  The DGP LHS 
performs better than the GP LHS due to the nonstationarity of the function.}

\begin{figure}[H]
	\includegraphics[width=0.45\linewidth]{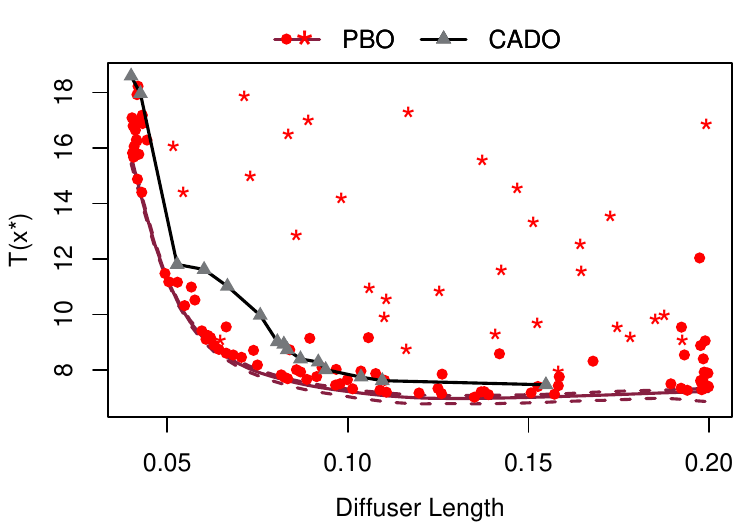}
	\caption{\textcolor{blue}{$\hat{T}(x^\star)$ (maroon) for the RDC diffuser simulation
	after a 50-point LHS (red stars) and 100 PBO acquisitions (red circles). 
	The black line and gray triangles indicate CADO's Pareto front.}}
	\label{fig:fulldiffuser}
\end{figure}

\textcolor{blue}{Finally, in Figure~\ref{fig:fulldiffuser} we include the 
estimated profile optima resulting
from all 100 acquisitions (150 total observations) of our PBO procedure on the 
RDC diffuser simulation.  We again overlay the Pareto front found by the
multi-objective optimization which used 113 observations.}

\section{Computation Time}\label{supp:time}

\textcolor{blue}{The computation times required for any of the design procedures
entertained in Section~\ref{sec:toyproblems} is significantly affected by $n$, 
$m$, $d$, the surrogate used, and the number of joint posterior samples used to 
estimate $\hat{T}(x^\star)$.  Traditional GP surrogates require far less computation
than DGP surrogates, which require inference of the latent warping.  The 
majority of computation is spent on generating joint posterior
samples at the modified tricands locations.  Even with Vecchia approximation to
avoid cubic costs, this computation can be hefty if the number of samples needed
and/or the number of candidate locations is very large.}

\textcolor{blue}{To provide further context, Table~\ref{tab:runtime} reports
the computation time required to complete an entire sequential design for
each synthetic example entertained in Section~\ref{sec:toyproblems}.  More computation
is required for larger $d$ and $m$, but the most notable increase occurs in the
switch from a GP to a DGP surrogate.  The time required to complete our PBO procedure
and the PEI procedure are nearly identical since both use the same modified tricands.}

\begin{table}[H]
	\textcolor{blue}{\begin{tabular}{|c|c|c|c|c|}
	\hline
	Function, Dimension & Runtime & Runtime & Number of & Surrogate \\
	\& Stationarity & PBO & PEI & Added Points & Used \\
	\hline
	2D stationary Branin & 1.05 min & 1.07 min & 20 & GP \\
	3D stationary Kyger & 3.07 min & 3.08 min & 30 & GP\\
	4D nonstationary squiggle & approx. 4.5 h & -- & 80 & DGP \\
	\hline
	\end{tabular}}
	\caption{\textcolor{blue}{Computation times required for a single run of
	PBO and PEI for the test functions of Section~\ref{sec:toyproblems}.}}
	\label{tab:runtime}
\end{table}

\textcolor{blue}{The previous runs used 1,000 joint posterior samples for each
estimation of $T(x^\star)$.  Computation time is largely driven by the number
of posterior samples needed, but reducing the number too drastically could affect
performance.  To investigate, we ran our PBO procedure on the 2D Branin function and
the 4D squiggle function while varying the number of posterior samples used in each
estimation of $T(x^\star)$.  Simulation settings otherwise match those described
in Section~\ref{sec:toyproblems}.  The resulting performance and computation times are
reported in Tables~\ref{tab:braninsamples} and \ref{tab:squigglesamples}.  While we
consider up to 5,000 samples when using the GP surrogate, this many samples was infeasible
for the DGP surrogate.}

\begin{table}[H]
	\textcolor{blue}{\begin{tabular}{|c|c|c|c|c|c|}
		\hline
		Number of Samples & Coverage & AvgCI & RMSE & MaxAD & Time \\
		\hline
		200 & 0.82 & 2.257 & 0.263 & 0.876 & 1.03 min \\
		1000 & 1 & 1.255 & 0.245 & 0.583 & 1.05 min\\
		5000 & 1 & 1.156 & 0.22 & 0.719 & 1.20 min \\
		\hline
		\end{tabular}}
	\caption{\textcolor{blue}{Results from PBO for the 2D Branin function
	 using a GP surrogate with $n=10$ and $m=30$, while varying the number of 
	 posterior samples used in each estimation of $T(x^\star)$.}}
	\label{tab:braninsamples}
\end{table}

\begin{table}[H]
	\textcolor{blue}{\begin{tabular}{|c|c|c|c|c|c|}
			\hline
			Number of Samples & Coverage & AvgCI & RMSE & MaxAD & Time \\
			\hline
			200 & 0.48 & 0.244 & 0.263 & 0.246 & About 35 min\\
			1000 & 0.2 & 0.106 & 0.077 & 0.127 & About 4.5 hrs\\
			\hline
	\end{tabular}}
	\caption{\textcolor{blue}{Results from PBO for the 4D squiggle function
	 using a DGP surrogate with $n=40$ and $m=120$, while varying the number of 
	 posterior samples used in each estimation of $T(x^\star)$.}}
	\label{tab:squigglesamples}
\end{table}

\textcolor{blue}{While coverage and accuracy generally benefit from the use 
of more samples, the computation time increases drastically in the nonstationary 
cases with the DGP surrogate, which disallows the use of more posterior 
samples to estimate $T(x^\star)$.}

\end{document}